\pdfoutput=1  

\documentclass[aps,twocolumn,groupedaddress,floatfix,preprintnumbers,nofootinbib,eqsecnum]{revtex4}

\usepackage{amsmath, float, graphicx,placeins,amssymb,multirow,pgfplots,pgfplotstable}
\usepackage{tikz,hyperref}
\usetikzlibrary{shapes.geometric,arrows}

\begin{document}
\title{
Lines and Boxes:\\  Unmasking Dynamical Dark Matter through Correlations\\ in the MeV Gamma-Ray Spectrum
}

\author{Kimberly K.\ Boddy$^{1}$\footnote{E-mail address:  {\tt kboddy@hawaii.edu}},
      Keith R.\ Dienes$^{2,3}$\footnote{E-mail address:  {\tt dienes@email.arizona.edu}},
      Doojin Kim$^{4}$\footnote{E-mail address:  {\tt immworry@ufl.edu}},\\
      Jason Kumar$^{1}$\footnote{E-mail address:  {\tt jkumar@hawaii.edu}},
      Jong-Chul Park$^{5}$\footnote{E-mail address:  {\tt jcpark@cnu.ac.kr}},
      Brooks Thomas$^{6,7}$\footnote{E-mail address:  {\tt thomasbd@lafayette.edu}}\\ ~}
\affiliation{
     $^1\,$Department of Physics \& Astronomy, University of Hawaii, Honolulu, HI 96822  USA\\
     $^2\,$Department of Physics, University of Arizona, Tucson, AZ  85721  USA\\
     $^3\,$Department of Physics, University of Maryland, College Park, MD  20742  USA\\
     $^4\,$Department of Physics, University of Florida, Gainesville, FL 32611  USA\\
     $^5\,$Department of Physics, Chungnam National University, Daejeon 34134  Korea\\
     $^6\,$Department of Physics, Colorado College, Colorado Springs, CO  80903  USA\\
     $^7\,$Department of Physics, Lafayette College, Easton, PA  18042  USA}

\begin{abstract}
Identifying signatures of dark matter at indirect-detection experiments is
generally more challenging for scenarios involving non-minimal dark sectors such as Dynamical
Dark Matter (DDM) than for scenarios involving a single dark particle.  This
additional difficulty arises because the partitioning of the total dark-matter
abundance across an ensemble of different constituent particles with different
masses tends to ``smear'' the injection spectra of photons and other cosmic-ray
particles that are produced via dark-matter annihilation or decay.  As a result,
the imprints of the dark sector on these cosmic-ray flux spectra typically
take the form of continuum features rather than sharp peaks or lines.  In this paper, however,
we identify an unambiguous signature of non-minimal dark sectors such as DDM
which can overcome these issues and potentially be observed at gamma-ray
telescopes operating in the MeV range.  We discuss the specific situations in
which this signature can arise, and demonstrate that this signature can be exploited in order to
significantly enhance our ability to resolve the unique spectral features of DDM and other non-minimal
dark sectors at future gamma-ray facilities.
\end{abstract}

\maketitle


\newcommand{\PRE}[1]{{#1}} 
\newcommand{\ul}{\underline}
\newcommand{\del}{\partial}
\newcommand{\nbox}{{\,\lower0.9pt\vbox{\hrule \hbox{\vrule height 0.2 cm
\hskip 0.2 cm \vrule height 0.2 cm}\hrule}\,}}

\newcommand{\postscript}[2]{\setlength{\epsfxsize}{#2\hsize}
   \centerline{\epsfbox{#1}}}
\newcommand{\gweak}{g_{\text{weak}}}
\newcommand{\mweak}{m_{\text{weak}}}
\newcommand{\mplanck}{M_{\text{Pl}}}
\newcommand{\mstar}{M_{*}}
\newcommand{\sigmaan}{\sigma_{\text{an}}}
\newcommand{\sigmatot}{\sigma_{\text{tot}}}
\newcommand{\sigmaSI}{\sigma_{\rm SI}}
\newcommand{\sigmaSD}{\sigma_{\rm SD}}
\newcommand{\OmegaM}{\Omega_{\text{M}}}
\newcommand{\OmegaDM}{\Omega_{\text{DM}}}
\newcommand{\ipb}{\text{pb}^{-1}}
\newcommand{\ifb}{\text{fb}^{-1}}
\newcommand{\iab}{\text{ab}^{-1}}
\newcommand{\ev}{\text{eV}}
\newcommand{\kev}{\text{keV}}
\newcommand{\mev}{\text{MeV}}
\newcommand{\gev}{\text{GeV}}
\newcommand{\tev}{\text{TeV}}
\newcommand{\pb}{\text{pb}}
\newcommand{\mb}{\text{mb}}
\newcommand{\cm}{\text{cm}}
\newcommand{\m}{\text{m}}
\newcommand{\km}{\text{km}}
\newcommand{\kg}{\text{kg}}
\newcommand{\g}{\text{g}}
\newcommand{\s}{\text{s}}
\newcommand{\yr}{\text{yr}}
\newcommand{\Mpc}{\text{Mpc}}
\newcommand{\etal}{{\em et al.}}
\newcommand{\eg}{{\em e.g.}}
\newcommand{\ie}{{\em i.e.}}
\newcommand{\ibid}{{\em ibid.}}
\newcommand{\Eqref}[1]{Equation~(\ref{#1})}
\newcommand{\secref}[1]{Sec.~\ref{sec:#1}}
\newcommand{\secsref}[2]{Secs.~\ref{sec:#1} and \ref{sec:#2}}
\newcommand{\Secref}[1]{Section~\ref{sec:#1}}
\newcommand{\appref}[1]{App.~\ref{sec:#1}}
\newcommand{\figref}[1]{Fig.~\ref{fig:#1}}
\newcommand{\figsref}[2]{Figs.~\ref{fig:#1} and \ref{fig:#2}}
\newcommand{\Figref}[1]{Figure~\ref{fig:#1}}
\newcommand{\tableref}[1]{Table~\ref{table:#1}}
\newcommand{\tablesref}[2]{Tables~\ref{table:#1} and \ref{table:#2}}
\newcommand{\Dsle}[1]{\slash\hskip -0.28 cm #1}
\newcommand{\met}{{\Dsle E_T}}
\newcommand{\mpt}{\not{\! p_T}}
\newcommand{\Dslp}[1]{\slash\hskip -0.23 cm #1}
\newcommand{\Dsl}[1]{\slash\hskip -0.20 cm #1}

\newcommand{\mB}{m_{B^1}}
\newcommand{\mq}{m_{q^1}}
\newcommand{\mf}{m_{f^1}}
\newcommand{\mKK}{m_{KK}}
\newcommand{\WIMP}{\text{WIMP}}
\newcommand{\SWIMP}{\text{SWIMP}}
\newcommand{\NLSP}{\text{NLSP}}
\newcommand{\LSP}{\text{LSP}}
\newcommand{\mWIMP}{m_{\WIMP}}
\newcommand{\mSWIMP}{m_{\SWIMP}}
\newcommand{\mNLSP}{m_{\NLSP}}
\newcommand{\mchi}{m_{\chi}}
\newcommand{\mgravitino}{m_{\gravitino}}
\newcommand{\mmed}{M_{\text{med}}}
\newcommand{\gravitino}{\tilde{G}}
\newcommand{\Bino}{\tilde{B}}
\newcommand{\photino}{\tilde{\gamma}}
\newcommand{\stau}{\tilde{\tau}}
\newcommand{\slepton}{\tilde{l}}
\newcommand{\snu}{\tilde{\nu}}
\newcommand{\squark}{\tilde{q}}
\newcommand{\mgaugino}{M_{1/2}}
\newcommand{\epsEM}{\varepsilon_{\text{EM}}}
\newcommand{\mmess}{M_{\text{mess}}}
\newcommand{\lmess}{\Lambda}
\newcommand{\nmess}{N_{\text{m}}}
\newcommand{\signmu}{\text{sign}(\mu)}
\newcommand{\Omegachi}{\Omega_{\chi}}
\newcommand{\lambdafs}{\lambda_{\text{FS}}}
\newcommand{\be}{\begin{equation}}
\newcommand{\ee}{\end{equation}}
\newcommand{\bea}{\begin{eqnarray}}
\newcommand{\eea}{\end{eqnarray}}
\newcommand{\beq}{\begin{equation}}
\newcommand{\eeq}{\end{equation}}
\newcommand{\beqn}{\begin{eqnarray}}
\newcommand{\eeqn}{\end{eqnarray}}
\newcommand{\baln}{\begin{align}}
\newcommand{\ealn}{\end{align}}
\newcommand{\lsim}{\lower.7ex\hbox{$\;\stackrel{\textstyle<}{\sim}\;$}}
\newcommand{\gsim}{\lower.7ex\hbox{$\;\stackrel{\textstyle>}{\sim}\;$}}

\newcommand{\ssection}[1]{{\em #1.\ }}
\newcommand{\rem}[1]{\textbf{#1}}

\def\ie{{\it i.e.}\/}
\def\eg{{\it e.g.}\/}
\def\etc{{\it etc}.\/}
\def\calN{{\cal N}}

\def\mptwo{{m_{\pi^0}^2}}
\def\mp{{m_{\pi^0}}}
\def\sqtsn{\sqrt{s_n}}
\def\sqtsn{\sqrt{s_n}}
\def\sqtsn{\sqrt{s_n}}
\def\sqts0{\sqrt{s_0}}
\def\Dsqts{\Delta(\sqrt{s})}
\def\Omegatot{\Omega_{\mathrm{tot}}}


\section{Introduction\label{sec:Introduction}}


Understanding the properties of the dark sector represents one of the 
great experimental and theoretical challenges facing physics today.
Indeed, we even lack insight into such fundamental questions as 
whether the dark sector
is minimal (\eg, consisting of only one or a few dark particle species)
or non-minimal (\eg, consisting of many particle species).
A pressing phenomenological question, therefore, 
is to determine how --- and 
to what degree it is even possible --- 
to experimentally distinguish non-minimal dark sectors 
from their more traditional, minimal counterparts.
This is especially true for scenarios within the
Dynamical Dark Matter (DDM)~\cite{DDM1,DDM2} framework --- a framework for
dark-matter physics in which the dark-matter ``candidate'' is an ensemble
consisting of a potentially vast number of individual constituent particle
species exhibiting a variety of masses, decay widths, and cosmological 
abundances.
Such DDM dark sectors give rise to 
collective phenomena that transcend 
expectations based on traditional dark-matter frameworks.
For example, the phenomenological viability of such a DDM ensemble as 
a representation of the dark sector 
rests not on the stability of each of these species
individually, but rather on a subtle balancing between decay
widths and cosmological abundances across the ensemble as a whole.  

In many DDM scenarios, the ensemble constituents share the same or similar quantum numbers.
In such cases, the detection channels through which one might hope to
find evidence of such an ensemble are essentially identical to those in which one
would seek evidence of a traditional dark-matter candidate with the
identical quantum numbers.  
However, even if the ensemble constituents share similar quantum numbers, they 
generically differ in their masses and couplings.
As a result,
it is often possible
to distinguish DDM ensembles and other non-minimal dark sectors experimentally
by analyzing the distributions of relevant kinematic variables.
At direct-detection experiments, for example, the relevant distribution
is the recoil-energy spectrum of the recoiling nucleus~\cite{DDMDD}.  
Likewise, at indirect-detection experiments, 
the relevant kinematic distributions are the differential flux spectra
of the SM particles which can be produced via dark-matter annihilation or decay~\cite{DDMAMS}.
Finally, at colliders,
the relevant distributions are those corresponding to  
a number of well-chosen kinematic variables formed from the momenta of Standard-Model (SM)
particles produced alongside the dark-matter particles.
The information contained in the full shapes of these distributions can be used to
distinguish DDM from traditional dark-matter scenarios~\cite{DDMLHC1,DDMLHC2},  
and indeed can be used to 
distinguish a variety of other non-minimal dark-matter scenarios as well~\cite{doojin1,doojin2,doojin3,doojin4}.

A variety of cosmic-ray particles --- among them electrons, positrons, photons,
antiprotons, neutrinos, and antideuterons --- can potentially yield information about
the nature of the dark matter.  For example, it has been shown that DDM ensembles can give
rise to characteristic signatures~\cite{DDMAMS} in the flux spectra of electrons and
positrons which can account for the positron excess observed by
PAMELA~\cite{PAMELA}, AMS-02~\cite{AMS}, and a
host of other experiments --- most notably without predicting an abrupt downturn in the positron
fraction at high energies.  However, of all cosmic-ray particles whose
flux spectra we have the ability to measure, photons are the particles that afford the greatest
potential for probing the structure of the dark sector.  
This is true primarily for two reasons.
First, the spectrum of
photons injected by a particular source is deformed
far less by interactions with the interstellar medium (ISM) than are the spectra associated
with most other cosmic-ray particles.  Thus, features imprinted on the photon spectrum at
injection --- features which might be indicative of dark-sector non-minimality --- are 
not washed out as a result
of their propagation through this medium.  Second, unlike neutrinos (which are also largely
unaffected by propagation through the ISM), photons are easy to detect and their energies
and directions can be measured with great precision.

It nevertheless remains true
that identifying such signatures at indirect-detection
experiments is generally more challenging for DDM scenarios
than for other, more traditional dark-matter scenarios.  This is because 
the injection spectra of photons and other cosmic-ray particles from
dark-matter annihilation or decay within DDM scenarios are subject to an additional ``smearing''
effect due to the partitioning of the total dark-matter abundance across an
ensemble of constituent particles with a range of masses.  Thus, the characteristic
imprints which these ensembles leave in the corresponding flux spectra 
typically take the form of continuum features rather than sharp peaks or lines.
This is especially true for cases in which the splittings between the masses of ensemble constituents are
small.  Disentangling continuum features from astrophysical backgrounds is
generally significantly more challenging than disentangling sharp peaks or lines.  
Moreover, even in situations in which
such features can be robustly identified, it is often impossible to conclusively determine 
whether dark matter or some more mundane astrophysical process is responsible.

In this paper, we identify an unambiguous signature of DDM (and of non-minimal dark
sectors more generally) which 
can serve to overcome these issues
and potentially be observed at gamma-ray telescopes 
sensitive to photons with energies in the $\mathcal{O}(1 - 100)~{\rm MeV}$ range.
This signature arises in cases in
which each of the ensemble constituents annihilates or decays predominately into
a primary photon and a neutral pion~\cite{Boddy:2015efa}, the latter subsequently decaying
into a pair of secondary photons.  

In general, the primary photons give rise to a line-like feature, while the
secondary photons give rise to a characteristic box-like feature whose width
is related to the energy (or boost) of the decaying pion.
(We review the kinematics of these processes in the Appendix.)
In the case of a single dark-matter species,
this combination of a line-like feature and a box-like feature
is notable and distinctive.
Such features have previously been studied,
 {\it e.g.}\/, in Refs.~\cite{Ibarra:2012dw, Boddy:2015efa, Garcia-Cely:2016pse}. 
In the case of a DDM ensemble, by contrast,
the primary photons give rise to a
{\it set}\/ of line-like features
while the secondary photons give rise to a {\it set}\/ of  box-like features.
In this paper, we are particularly interested in the regime in which the splitting 
between constituent masses is small compared to the energy resolution of the telescope.
In such cases, the set of line-like features will appear a single effective {\it continuum}\/ spectral feature.
Likewise, the pion energies will also form an effective continuum which then produces a continuum of box-like features. 
Note that in this context the pion energies will form an effective  continuum because these pions are produced via 
the direct annihilation or decay 
of the different DDM ensemble components which themselves exhibit an effective continuum of masses.
This is therefore somewhat different than the continuous pion spectra 
that might emerge through multiple sequential decays, 
as in  Refs.~\cite{Kim:2015usa,Kim:2015gka}, 
or via $n$-body decays with $n\geq 3$.

Taken in isolation, each of these two spectral features 
reveals information about the properties of the DDM ensemble.  
However, what makes this
signature particularly advantageous from the perspective of distinguishing
between minimal and non-minimal DDM dark sectors is that the spectral shapes of these two
features are {\it correlated}\/.  Thus, a comparison between the information independently extracted
from these two continuum features can provide a powerful consistency check that
they indeed have a common underlying origin in terms of a DDM ensemble.  
Indeed, it was shown in Ref.~\cite{Boddy:2015efa} that for a single-particle dark-matter candidate
which decays into this same final state, correlations between the properties of
the line and box features in the gamma-ray spectrum could be used to reconstruct the
mass of the dark-matter particle.   By contrast, in a DDM context, we shall see that the correlations
between the corresponding continuum features can be used to reconstruct the fundamental
relations which describe how the the masses, abundances, and lifetimes of the
ensemble constituents scale across the ensemble as a whole.

This paper is organized as follows.  In Sect.~\ref{sec:model},
we discuss the circumstances under which the constituents of a DDM ensemble
annihilate or decay predominately to a $\gamma \pi^0$ final state.
We also  establish the conventions we shall use for parametrizing such an ensemble.
In Sect.~\ref{sec:spectrum}, we then calculate the contribution to the differential
photon flux which arises from dark-matter annihilation or decay in such scenarios.
We also discuss the two distinctive features which arise in the flux spectrum and
examine how the spectral shapes of these features, and the degree to which they
overlap, vary as a function of the parameters which characterize the ensemble.
In Sect.~\ref{sec:prospects}, we investigate the prospects of identifying these spectral features 
in the diffuse galactic gamma-ray spectrum and in the gamma-ray spectra
of dwarf spheroidal galaxies at the next generation of gamma-ray telescopes,
and in Sect.~\ref{sec:measurement} we examine
the degree to which the underlying parameters which characterize the DDM ensemble
can be extracted from the spectral shapes of these features.  
Finally, in Sect.~\ref{sec:conclusions},
we summarize our conclusions 
and provide an outlook for future work,
while in the Appendix we review the kinematics 
leading to line-like and box-like features in the photon spectrum.


\section{DDM Ensembles and Their Decays to Photons and Pions\label{sec:model}}


Within the context of DDM framework~\cite{DDM1,DDM2}, the dark sector comprises a potentially vast ensemble
of individual particle species $\phi_n$ whose cosmological abundances $\Omega_n$
are balanced against their decay widths $\Gamma_n$ in such a way as to ensure
consistency with observational data.
It turns out that DDM ensembles arise naturally in a variety
of well-motivated extensions of the SM;  these include
scenarios which involve extra
spacetime dimensions~\cite{DDM1,DDM2,DDMAxion}, 
large spontaneously-broken symmetry groups~\cite{RandomMatrixDDM},  
confining hidden-sector gauge groups~\cite{HagedornDDM}, 
or bulk physics in open string theories~\cite{HagedornDDM,bhupal}. 
In what follows, we adopt the convention
that the index $n= 0, 1, 2, \ldots, N$ labels the
particles in order of increasing mass.

Our principal aim in this paper is to study
the astrophysical gamma-ray signatures associated with DDM ensembles in which the
ensemble constituents annihilate or decay predominately into a $\gamma \pi^0$
final state (with a subsequent pion decay $\pi^0\to \gamma\gamma$), 
and to determine the degree to which information about
the ensemble can be extracted from these signatures.
Such final states can arise in 
DDM scenarios in which the $\phi_n$ couple directly to quarks via an effective contact
operator~\cite{Boddy:2015efa}.  The structure of this operator can be inferred from the
fact that the final state $\gamma \pi^0$ is odd under charge-conjugation.  
Under the assumption that the $SU(2)$ weak interaction can be neglected and that the fundamental
interactions between the ensemble constituents and the SM fields are $C$-invariant, the
initial state must therefore be $C$-odd as well.  One possible operator
structure which possesses the appropriate symmetry properties is
\begin{equation}
  \mathcal{O}_n ~=~ c_n \, B^\mu_n\,  \bar q \gamma_\mu q~
\end{equation}
where $c_n$ is an operator coefficient and where $B^\mu_n$ is a $C$-odd
quantity involving the $\phi_n$ fields alone.  One situation in which
an operator of this sort arises is that in which the $\phi_n$ are spin-1 fields $\phi_n^\mu$
and corresponds to the case in which $B^\mu_n$ is identified with the field $\phi_n^\mu$
itself.  In this case, the operator gives rise to decay processes of the form
$\phi_n \rightarrow \gamma \pi^0$.  Another situation in which such an operator
arises is that in which $B^\mu_n = \mathcal{J}_n^\mu / \Lambda^2$, where
$\mathcal{J}_n^\mu$ is an approximately conserved current associated with the 
particle number of the ensemble constituent $\phi_n$.  In this case,
the operator gives rise to annihilation processes of the form
$\phi_n^\dagger \phi_n \rightarrow \gamma \pi^0$.  In both of these cases,
the fundamental interaction between the 
dark ensemble constituents $\phi_n$ and SM quarks gives rise to an
effective operator of the form~\cite{Boddy:2015efa}
\begin{equation}
  \mathcal{O}_{n,\textrm{eff}} ~=~ \frac{e \,c_n}{16 \pi^2 f_\pi }
  B_{\mu,n}\,  F_{\nu\rho}\,
   (\partial_{\sigma}\pi^0)\, \epsilon^{\mu\nu\rho\sigma}~
\end{equation}
in the low-energy confined phase of the theory.

We have shown that there exists a self-consistent mechanism through which the
constituents of a DDM ensemble can be coupled to the photon and neutral-pion fields.
However, whether or not processes resulting in a $\gamma \pi^0$ final state dominate
the decay width or annihilation cross-section of a given $\phi_n$ also depends
on the center-of-mass (CM) energy $\sqtsn$ associated with those processes.  Since a
number of considerations imply that the velocities of dark-matter particles
within the halos of galaxies are non-relativistic, the CM energy for the
annihilation or decay of an ensemble constituent with mass $m_n$ is well
approximated by
\begin{equation}
  \sqtsn ~\approx~ \begin{cases}
    2m_n & \mbox{for annihilation}\\
    m_n & \mbox{for decay}~.
  \end{cases}
\label{sndef}
\end{equation}
Moreover, the assumption that the dark matter is non-relativistic also implies
that the CM frame for annihilating/decaying dark-matter particles is effectively
equivalent to the rest frame of the instrument which detects the annihilation/decay
products.

In the regime in which $\sqtsn < m_{\pi^0}$,
annihilation/decay to a photon and an on-shell $\pi^0$ is kinematically forbidden.
Annihilation/decay to a three-photon final state can still proceed in this regime
via an off-shell $\pi^0$, but processes of this sort do not give rise to the
same characteristic features in the photon spectrum.  On the other hand, in the
regime in which  $\sqtsn > 2m_{\pi^\pm}$, the annihilation/decay of $\phi_n$
to $\pi^+\pi^-$ is kinematically allowed, but the photons produced as 
final-state radiation in conjunction with charged-pion production can contribute significantly to
the photon flux and overwhelm the contribution from $\gamma \pi^0$.  Thus, 
the range of CM energies for which
the $\gamma \pi^0$ channel provides the dominant contribution to the photon flux is given by
\begin{equation}
          m_{\pi^0} < \sqrt{s_n} < 2 m_{\pi^\pm}~,
\label{snrange}
\end{equation}
corresponding to the dark-matter mass ranges
\begin{equation}
             \begin{cases}
     {\textstyle{1\over 2}} m_{\pi^0} < m_n < m_{\pi^\pm} &  \mbox{for annihilation}\\
            m_{\pi^0} < m_n < 2m_{\pi^\pm} & \mbox{for decay}~.
           \end{cases}
\end{equation}
For simplicity, in what follows we shall focus on DDM ensembles in which the masses
of all of the ensemble constituents lie within this range.
We are therefore interested
in DDM ensembles in which the mass scale of the $\phi_n$ is of order
$m_n \sim \mathcal{O}(100)~{\rm MeV}$.~  
Indeed, the collective contribution to the photon flux from the annihilation/decay of 
any lighter constituents in the DDM ensemble is typically negligible unless the density of such states is enormous.

For an ensemble constituent within our chosen mass range, the $\gamma \pi^0$
channel generically yields the dominant contribution to the photon flux.  The only other
two-body final states which are consistent with the symmetries
of the theory and kinematically accessible within the range in Eq.~(\ref{snrange}) 
are $\bar{\nu}\nu$, $e^+e^-$, and $\mu^+\mu^-$.  The first of these is
irrelevant for photoproduction, while the contribution to the photon flux from
the other two are necessarily suppressed by additional factors of either $\alpha$ or 
$s_n G_F$, where $G_F$ is the Fermi constant.  
Consequently these processes will be comparatively insignificant
whenever the $\gamma \pi^0$ state is accessible.  The contributions associated with
final states involving three or more SM particles are likewise suppressed.

In general, the underlying mass spectrum of our DDM ensemble depends on
the type of ensemble under study, and as such it can be arbitrary.
For concreteness, however, we shall focus on the case in
which the mass spectrum of our DDM ensemble 
takes the generic form 
\begin{equation}
  m_n ~=~ m_0 + n^{\delta}\Delta m~
\label{mnspectrum}
\end{equation}
where $m_0$ is the mass of $\phi_0$ (the lightest of the $\phi_n$) and where the
mass splitting $\Delta m$ and scaling exponent $\delta$ are free parameters describing our
underlying DDM ensemble.
Indeed, many realistic DDM ensembles have mass spectra which follow exactly this generic
form.
Thus, the spectrum of corresponding CM energies takes the form
\begin{equation}
  \sqrt{s_n} ~=~ \sqrt{s_0} + n^{\delta}\Delta(\sqrt{s})~
\label{snspectrum}
\end{equation}
where $\Delta(\sqrt{s}) = \Delta m$ for decay  
and $\Delta(\sqrt{s}) = 2\Delta m$ for annihilation.
The splittings $\Delta(\sqtsn) \equiv \sqrt{s_{n+1}}- \sqtsn$ between the 
CM energies for the annihilation/decay of adjacent ensemble states 
are therefore given by
\begin{equation}
 \Delta(\sqrt{s_n})= \left[(n+1)^{\delta}-n^{\delta} \right]\Delta (\sqrt{s})~. 
\end{equation}
The case with $\delta=1$ is particularly interesting, occurring
when $\phi_n$ are the modes in
a Kaluza-Klein tower.  We shall therefore focus on this case in what follows.
For this value of $\delta$, the mass splitting
$m_{n+1}-m_n$ is uniform across the ensemble,
and $\Delta(\sqtsn) \equiv \Dsqts$ for all $n$.


\section{Gamma-Ray Spectrum from DDM Annihilations/Decays\label{sec:spectrum}}


In this section, we examine the signal contribution to the differential photon flux
$d\Phi/dE_\gamma$ which arises in DDM scenarios in which the ensemble constituents
annihilate/decay to a $\gamma \pi^0$ final state (thereby producing a single
``primary'' photon), followed by a subsequent
decay $\phi^0\to \gamma\gamma$ (thereby producing two ``secondary'' photons).    We begin with a
derivation of the general expression for this signal contribution, followed by
a discussion of the distinctive qualitative features in the flux spectrum which
arise in these scenarios.
Note that the kinematics of the $\phi_n\to \gamma \pi^0\to \gamma\gamma\gamma$ process
is reviewed in the Appendix.

\subsection{Differential photon flux: Quantitative results}

In order to derive an expression for the total differential photon flux $d\Phi_n/dE_\gamma$ coming
from anniliation and/or decay of the DDM ensemble, we begin by deriving 
an expression for the photon flux $\Phi_n$ coming from each individual ensemble constituent.
This is not particularly difficult, as there are only two primary ingredients that enter
into such a calculation.
The first is the integrated
energy density $\rho_n$  
(or squared energy density $\rho_n^2$)
of the $\phi_n$ component
along the line of sight:
\begin{equation}
  J_n ~\equiv~ \int d\Omega \int_{\textrm{LOS}}d\ell \times\begin{cases}
    \begin{array}{ll}
      \rho_n^2 & \textrm{ for annihilation} \\
      \rho_n & \textrm{ for decay~,}
    \end{array} \end{cases}
\end{equation}
where the differential solid angle $d\Omega$ corresponds to our region of interest on the sky.
The second ingredient, by contrast, is the 
annihilation/decay rate $R_n$ of this component into photons:      
the decay rate for the $\phi_n$ component is nothing but $\Gamma_n$, while 
the annihilation rate is given by 
$\langle \sigma_n v \rangle/4m_n$
where
$\langle \sigma_n v \rangle$
is the thermally-averaged cross section
for the annihilation process $\phi^\dagger_n\phi_n \rightarrow \gamma \pi^0$.  
Putting the pieces together, the resulting photon flux is then given by
$\Phi_n =   \calN_n  (J_n/4\pi) (R_n/m_n)$, where 
$\calN_n\equiv \calN_n^{(p)} + \calN_n^{(s)}=3$
is the total number of primary plus secondary photons produced via the annihilation/decay
of each $\phi_n$.
 
{\it A priori}\/, 
it is difficult to determine the individual line-of-sight 
integrals $J_n$.
However, 
it is natural to suppose that the energy densities $\rho_n$ of the
individual $\phi_n$ within the galactic halo and within the halos of other
galaxies are proportional to their overall cosmological abundances.
In  other words, we shall assume that
$\rho_n/\rho_{\mathrm{tot}} = \Omega_n/\Omegatot$,
where $\rho_{\mathrm{tot}} = \sum_{n=0}^N \rho_n$.
Under this assumption, we can then 
define an overall $n$-independent ``$J$-factor'' which represents the {\it total}\/ energy density
integrated along the line of sight,
\begin{equation}
  J ~\equiv~ \int d\Omega \int_{\textrm{LOS}}d\ell \times\begin{cases}
    \begin{array}{ll}
      \rho_{\rm tot}^2 & \textrm{ for annihilation} \\
      \rho_{\rm tot} & \textrm{ for decay~,}
    \end{array} \end{cases}
\label{Jdef}
\end{equation}
whereupon our resulting photon flux $\Phi_n$ takes the general form
\begin{equation}
  \Phi_n ~=~ \calN_n  \, \frac{J}{4\pi} \, \frac{\Omega_n}{\Omegatot}\, \frac{\lambda_n}{m_n}~
  \label{eq:PhintotExp}
\end{equation}
with
\begin{equation}
  \lambda_n ~\equiv ~ \begin{cases} \displaystyle
    \frac{\Omega_n}{\Omega_\textrm{tot}}
      \frac{\langle \sigma_n v \rangle}{4m_n} & \textrm{ for annihilation} \\
    \rule[0pt]{0pt}{14pt}\Gamma_n & \textrm{ for decay~.}
  \end{cases}
\end{equation}
For simplicity, we assume that the cross-section for {\it co}\/-annihilation processes
of the form $\phi^\dagger_m\phi_n \rightarrow \gamma \pi^0$ with $m \neq n$
is negligible.

Given the result for the individual flux $\Phi_n$ in Eq.~(\ref{eq:PhintotExp}),
we can now derive the collective contribution to the {\it differential}\/ photon flux from the
annihilation/decay of the {\it entire}\/ DDM ensemble.  Indeed, this is nothing but 
the sum over the individual
contributions $d\Phi_n/dE_\gamma$ from each of the $\phi_n$:
\begin{eqnarray}
  \frac{d\Phi}{dE_\gamma} ~=~ \sum_{n=0}^N \frac{d\Phi_n}{dE_\gamma}  &=& 
    \sum_{n=0}^N \frac{J}{4\pi} \frac{\Omega_n}{\Omega_\textrm{tot}}
    \frac{\lambda_n}{m_n} \frac{d\mathcal{N}_{n}}{dE_\gamma}\nonumber\\
     &=&  \frac{\Phi_0}{3} \, \sum_{n=0}^N \frac{\Omega_n}{\Omega_0}
                          \frac{m_0}{m_n}
                                \frac{\lambda_n}{\lambda_0}
    \frac{d\mathcal{N}_{n}}{dE_\gamma}~,~~~~~~
\label{eq:DiffPhotonFlux}
\end{eqnarray}
where 
\begin{equation}
  \frac{d\mathcal{N}_{n}}{dE_\gamma} ~=~
    \frac{d\mathcal{N}_{n}^{({p})}}{dE_\gamma}
    +\frac{d\mathcal{N}_{n}^{({s})}}{dE_\gamma}~
\end{equation}
represents the differential number of photons
per unit $E_\gamma$ produced via a single annihilation/decay event involving
the constituent $\phi_n$.

Given the expression in Eq.~(\ref{eq:DiffPhotonFlux}), our next step 
is to understand how $\Omega_n$, $\lambda_n$, and $m_n$ depend on $n$.
For an arbitrary collection of dark-sector species, 
these quantities might not exhibit any regular behavior as functions of $n$.
In a DDM ensemble, however, the abundances, decay widths, and cross-sections of the 
different components 
all exhibit specific scaling relations 
as functions of $m_n$  
across the DDM ensemble. 
Indeed, such scaling relations (whether exact or approximate) 
tend to emerge naturally as a result of the various theoretical
structures underlying these ensembles. 
Of course,
since a gamma-ray telescope is at best only capable of measuring the differential photon
  flux $d\Phi/dE_\gamma$, we see 
from Eq.~(\ref{eq:DiffPhotonFlux})
that such an instrument is not sensitive 
to the individual scaling behaviors 
of these different quantities;   rather, it is only sensitive to 
the scaling behavior of the particular combination $\Phi_n\propto \Omega_n \lambda_n/m_n$.
Accordingly, 
  for concreteness, 
  we shall assume that 
  the fluxes $\Phi_n$ scale with $m_n$ according to a single power law of the form
  \begin{equation}
    \Phi_n ~=~ \Phi_0 \left(\frac{m_n}{m_0}\right)^\xi ~=~
       \Phi_0 \left(\frac{\sqtsn}{\sqts0}\right)^\xi~
    \label{eq:flux}
  \end{equation}
  where the masses/CM energies follow Eqs.~(\ref{mnspectrum}) and (\ref{snspectrum})
and where the exponent $\xi$ is taken to be a free parameter.
Indeed, this is tantamount to assuming that
\begin{equation}
             \frac{\Omega_n}{\Omega_0}
               \frac{\lambda_n}{\lambda_0} ~=~ \left( \frac{m_n}{m_0}\right)^{\xi+1}~=~ 
               \left( \frac{\sqtsn}{\sqts0}\right)^{\xi+1}~.
\end{equation}
  As such, the exponent $\xi$ reflects the internal theoretical structure of 
the DDM ensemble under study.
  Note that this parametrization 
is applicable to both annihilation and decay,
  although in general we expect the actual value of $\xi$ for the case of annihilation
  to differ from that for decay.

This parametrization allows us to recast
our expression for the differential photon flux in
Eq.~(\ref{eq:DiffPhotonFlux}) 
into the relatively simpler form
\begin{equation}
  \frac{d\Phi}{dE_\gamma} ~=~ \frac{\Phi_0}{3} \sum_{n=0}^N
    \left(\frac{\sqrt{s_n}}{\sqrt{s_0}}\right)^\xi
    \frac{d\mathcal{N}_{n}}{dE_\gamma}~.
  \label{eq:totalflux}
\end{equation}
Moreover, as discussed in the Introduction, we are primarily interested in
the regime for which $\Delta m \ll \Delta E_\gamma$ over the energy range
of interest, where $\Delta E_\gamma$ is the energy resolution of the detector.
Thus, since we expect $\Delta E_\gamma \lesssim E_\gamma \leq \sqrt{s_N}$, we
shall focus on the case in which $\Delta m \ll m_0$ and the sum over $n$
in Eq.~(\ref{eq:totalflux}) is well approximated by an integral over the
continuous variable $\sqrt{s}$.
We then obtain
\begin{equation}
  \frac{d\Phi}{dE_\gamma} ~\approx~ \frac{\Phi_0}{3\Delta(\sqrt{s})}
    \int_{\sqrt{s_0}}^{\sqrt{s_N}} d\sqrt{s}
    \left(\frac{\sqrt{s}}{\sqrt{s_0}}\right)^\xi
  \frac{d\mathcal{N}}{dE_\gamma}~,
  \label{eq:masterflux}
\end{equation}
where $\Delta(\sqrt{s})$ is defined in Eq.~(\ref{snspectrum})
and where $d\mathcal{N}/dE_\gamma$ is the differential number of photons per
unit $E_\gamma$ 
resulting from an  ensemble constituent annihilating or decaying with CM energy $\sqrt{s}$
into a $\gamma \pi^0 $ final state,
followed by a subsequent decay $\pi^0\to \gamma\gamma$.
Note that the integral in Eq.~(\ref{eq:masterflux}) continues to represent a sum
over ensemble constituents, with the contribution from any $\sqrt{s}$ representing
the contribution from that ensemble constituent which annihilates or decays with CM energy $\sqrt{s}$. 

Proceeding further requires knowledge of 
$d\mathcal{N}/dE_\gamma$.   However, this quantity includes contributions from both primary and secondary photons,
and these two classes of photons 
have very different kinematic features.
We shall therefore consider each of these classes separately.

As discussed in the Appendix,
the primary photons are all monochromatic, occupying a ``line'' with energy
\beq
 E_{\rm line} ~=~    \frac{ s-\mptwo}{2 \sqrt{s}}~.
\label{Eline}
\eeq
There is also only one such photon per constituent decay/annihilation. 
Thus the primary photon contribution to $d\mathcal{N}/dE_\gamma$
is simply
\beq
\frac{d\mathcal{N}^{(p)}}{dE_\gamma} ~=~ \delta(E_\gamma-E_{\rm line})~,
\eeq
whereupon the corresponding contribution to the flux in Eq.~(\ref{eq:masterflux}) is given by
\begin{eqnarray}
  \frac{d\Phi^{(p)}}{dE_\gamma} &\approx& \frac{\Phi_0}{3\Delta(\sqrt{s})}\,
      \left(\frac{\sqrt{s_*}}{\sqrt{s_0}}\right)^\xi
      \frac{2s_\ast}{s_\ast+m_{\pi^0}^2} \nonumber \\
    &&  ~~~~\times \,
      \Theta(\sqrt{s_\ast}-\sqrt{s_0})\,
      \Theta(\sqrt{s_N} -\sqrt{s_\ast})~~~~~~~~~~
  \label{eq:PhiP}
\end{eqnarray}
where $\Theta(x)$ is the Heaviside function and where
\begin{equation}
  \sqrt{s_\ast} ~\equiv~ \sqrt{E_\gamma^2 + m_{\pi^0}^2} + E_\gamma~.
  \label{eq:sqrtsn}
\end{equation}
Physically, this means that there is only one DM constituent whose decay or annihilation
contributes to the primary photon flux at any energy $E_\gamma$:  this is the constituent 
whose decay or annihilation
occurs with the CM energy $\sqrt{s_\ast}$ given in Eq.~(\ref{eq:sqrtsn}).

The secondary photons have a different kinematics, however.
As discussed in the Appendix, the secondary photons 
emerging from an ensemble constituent decaying or annihilating with CM energy $\sqrt{s}$
have energies which uniformly populate a ``box'' 
whose lower and upper limits are respectively given by
\beq
       E_{\rm box}^{-} ~=~  \frac{\mptwo}{2\sqrt{s}} ~,~~~~~~~ 
       E_{\rm box}^{+} ~=~  \frac{\sqrt{s}}{2} ~.~
\eeq
Moreover, there are two secondary photons from each such event. 
Thus, the normalized contribution from the secondary photons to the differential photon number per
unit $E_\gamma$ 
is given by
\beqn
   \frac{d\mathcal{N}^{(s)}}{dE_\gamma}
    &=& 2 \, 
    \frac
    {\Theta(E_\gamma - E_{\rm box}^-) \, 
        \Theta(E_{\rm box}^+ - E_\gamma )}
       {  E_{\rm box}^+ - E_{\rm box}^- } ~\nonumber\\ 
  &=&   \frac{4 \sqrt{s}}{s-\mptwo}\,
    {\Theta(E_\gamma - E_{\rm box}^-) \, 
        \Theta(E_{\rm box}^+ - E_\gamma )}~,\nonumber\\
\label{twothetas}
\eeqn
whereupon the corresponding secondary photon flux becomes
\beq
  \frac{d\Phi^{(s)}}{dE_\gamma} ~\approx~ 
         \frac{4\Phi_0}{3\Delta(\sqrt{s})}\,
    \int_{\sqrt{s_{\rm min}}}^{\sqrt{s_{N}}} d\sqrt{s}
    \left(\frac{\sqrt{s}}{\sqrt{s_0}}\right)^\xi
     \frac{\sqrt{s}}{s-\mptwo}
\label{secondint}
\eeq
with
\beq
     \sqrt{s_{\rm min}} ~\equiv~ {\rm min} \left[ \sqrt{s_N}, \, 
              {\rm max}\left(  \sqrt{s_0},  \, 2E_\gamma, \, \frac{\mptwo}{2E_\gamma}\right)\right]~.
\eeq
Indeed, for any given value of $E_\gamma$, the Heaviside 
theta-functions in Eq.~(\ref{twothetas}) restrict the 
values of $\sqrt{s}$ which contribute in Eq.~(\ref{secondint})
to those 
which are compatible not only with our original 
constraints $\sqrt{s_0}\leq \sqrt{s}\leq \sqrt{s_N}$ but also 
with the simultaneous constraints
$E_\gamma < E_{\rm box}^+$ (which requires $\sqrt{s} > 2E_\gamma$)  
and $E_\gamma > E_{\rm box}^-$ (which requires $\sqrt{s} > \mptwo/2E_\gamma$).
The result in Eq.~(\ref{secondint})
can then be integrated in closed form, yielding
\beqn
  \frac{d\Phi^{(s)}}{dE_\gamma} 
  &\approx& \frac{2\Phi_0}{3\Delta(\sqrt{s})} 
      \left( \frac{\mp}{\sqrt{s_0}} \right)^\xi \, \times\nonumber\\
  && ~~~ \left\lbrack
         B_{z_1} (-\xi/2,0)
        -B_{z_2} (-\xi/2,0)
        \right\rbrack~~~~~  
\label{eq:PhiS}
\eeqn
where $B_z(a,b)$ is the incomplete Euler beta-function,
with $z_1\equiv \mptwo/s_{\rm min}$ and $z_2\equiv \mptwo/s_N$.

In summary, the overall signal contribution to the differential photon flux in
DDM scenarios of this sort is the sum of the two individual contributions from
primary and secondary photons given in Eqs.~(\ref{eq:PhiP}) and~(\ref{eq:PhiS}),
respectively.

\subsection{Differential photon flux: Qualitative features\label{sec:case-studies}}

The spectral feature associated with primary photons, which is described by
Eq.~(\ref{eq:PhiP}), extends between 
$E_{\rm line}(\sqrt{s_0})$ and $E_{\rm line}(\sqrt{s_N})$.
The shape
of this feature is in large part dictated by the value of the index $\xi$.
However, since we are focusing on ensembles in which the CM energy
for the annihilation/decay of each of the $\phi_n$ falls
within the range $m_{\pi^0} < \sqrt{s_n} < 2m_{\pi^\pm}$, this feature typically
appears reasonably flat (unless the value of $\xi$ is extreme) and exhibits
a sharp cutoff at $E_\gamma = E_{\rm line}(\sqrt{s_N})$.

By contrast, the spectral feature associated with secondary photons, which is
described by Eq.~(\ref{eq:PhiS}), has a markedly different shape.
As discussed above, the individual contribution to $d\Phi^{(\mathrm{s})}/dE_\gamma$
from each $\phi_n$ consists of a flat, box-like feature centered at $E_\gamma = m_{\pi^0}/2$ on a logarithmic scale.  Thus, the total contribution to the secondary photon
flux consists of a ``tower'' of such boxes centered at this same value of
$E_\gamma$.  Since the width of each box is given by $(s_n-\mptwo)/2\sqrt{s_n}$,
the narrowest box is associated with the lightest ensemble constituent participating 
in the relevant annihilation/decay process, and has a
width $(s_0-m_{\pi^0}^2)/2\sqrt{s_0}$ if $\phi_0$ is indeed that constituent.  
This implies that in cases in
which $\sqrt{s_0} \approx m_{\pi^0}$, a sharp peak or spike appears at the center of
the tower~\cite{Stecker,Agashe:2012bn}.  By contrast, in cases in which the difference between $\sqrt{s_0}$
and $m_{\pi^0}$ is larger --- even by a few MeV --- the top of 
the tower appears flat and forms a plateau~\cite{Stecker,Agashe:2012bn,Kim:2015gka,Chen:2014oha}.
We thus have
\beq
 \begin{cases}
     \sqrt{s_0} \approx  \mp &  {\rm spike} \\ 
     \sqrt{s_0} >  \mp &  {\rm plateau}~. 
 \end{cases}
\label{peakcases}
\eeq

Another important consideration is whether and to what extent the spectral features
associated with primary and secondary photons in this scenario overlap.  Indeed,
as we shall see in Sect.~\ref{sec:measurement}, the degree of overlap between these
spectral features determines the fitting procedure which must be used in extracting
information about the fundamental parameters governing the DDM ensemble.
In particular, in cases in which the two features are well separated, a parametric
fit can be performed for each in isolation.  By contrast, in cases
in which the overlap between the two features is significant, a single fit must be
performed on the combined spectrum in order to extract the underlying
parameters governing the DDM ensemble.  In either case, however, we shall find
that it is often possible to measure most of the underlying
parameters which characterize the DDM ensemble with reasonable precision.

In order to assess the degree of overlap between the primary  and secondary photon
spectra for any particular choice of parameters, we compare the maximum possible
energy for a primary photon to the minimum possible energy for a secondary photon.
The former is given by
$E_{\rm line}(\sqrt{s_N})$ while
the latter is given by
$E_{\rm box}^-(\sqrt{s_N})$.
The spectral features associated with the primary and secondary photons
will thus overlap only if
$E_{\rm box}^-(\sqrt{s_N}) < E_{\rm line}(\sqrt{s_N})$,
or equivalently if 
$\sqrt{s_N} > \sqrt{2}\mp$.
We thus have
\beq
 \begin{cases}
     \sqrt{s_N} <  \sqrt{2} \mp &  {\rm no~overlap} \\ 
     \sqrt{s_N} >  \sqrt{2} \mp  &  {\rm overlap}~. 
 \end{cases}
\label{overlapcases}
\eeq

\begin{table*}[t]
\centering
\begin{tabular}{||c||c|c|c|c|c||}
\hline
\hline
~Benchmark~ &   ~$
\rule[-5pt]{0pt}{16pt}
\sqrt{s_0}$~(MeV)~ &  ~$\sqrt{s_N}$~(MeV)~ & ~$N$~ &   
            ~Behavior at \mbox{$E_\gamma=m_{\pi^0}/2$} ~ & ~Spectral overlap~ \\
\hline
A & 135  & 181  & 23  & spike & negligible \\
B & 135  & 231  & ~~48~~ & spike & significant \\
C & 164  & 180  & 8   & plateau & negligible \\
D & 164  & 230  & 33  & plateau & significant \\
\hline\hline
\end{tabular}
\caption{Four benchmark DDM ensembles --- each corresponding to a different
  choice of the parameters $\sqrt{s_0}$ and $\sqrt{s_N}$ ---  which illustrate the range of spectral
  signatures which arise in this scenario.  For each of these benchmarks, we have taken
    $\Delta (\sqrt{s}) = 2$~MeV.~  The resulting features (spike versus plateau at $E_\gamma = \mp/2$ 
   and the degree of spectral overlap) are governed by the criteria in Eqs.~(\ref{peakcases}) and (\ref{overlapcases}). 
  \label{tab:case}}
\end{table*}

\begin{figure*}[t]
\centering
\includegraphics[width=0.4\textwidth, keepaspectratio]{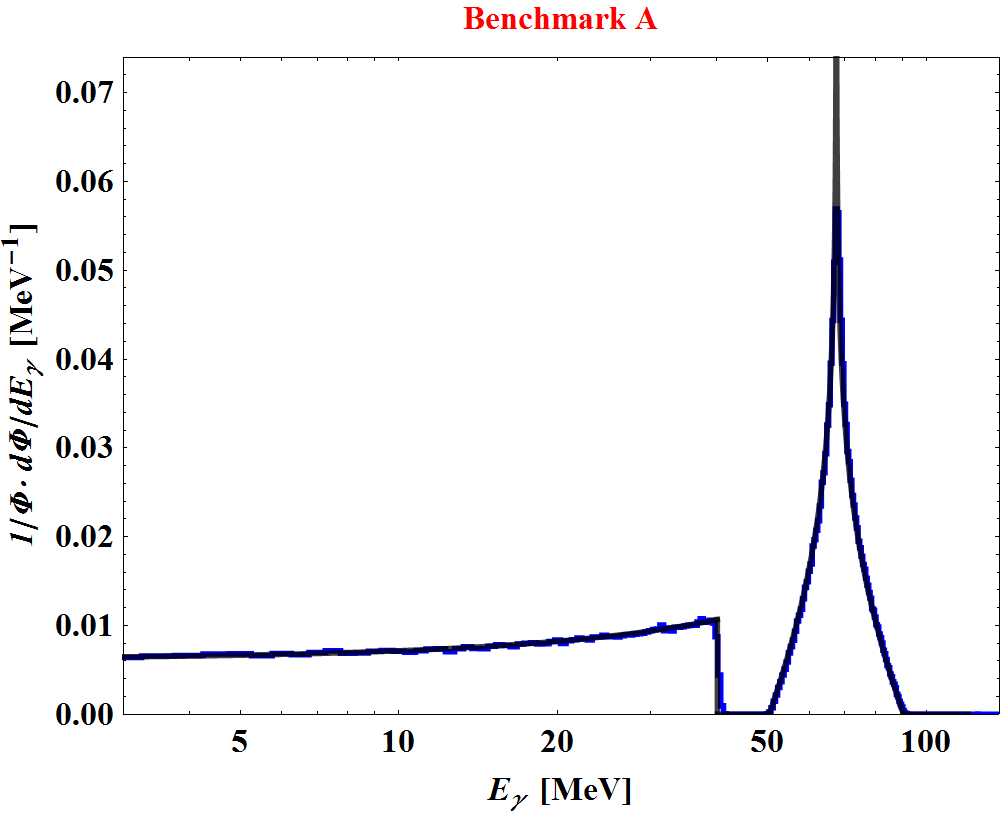}\hspace{1.2cm}
\includegraphics[width=0.4\textwidth, keepaspectratio]{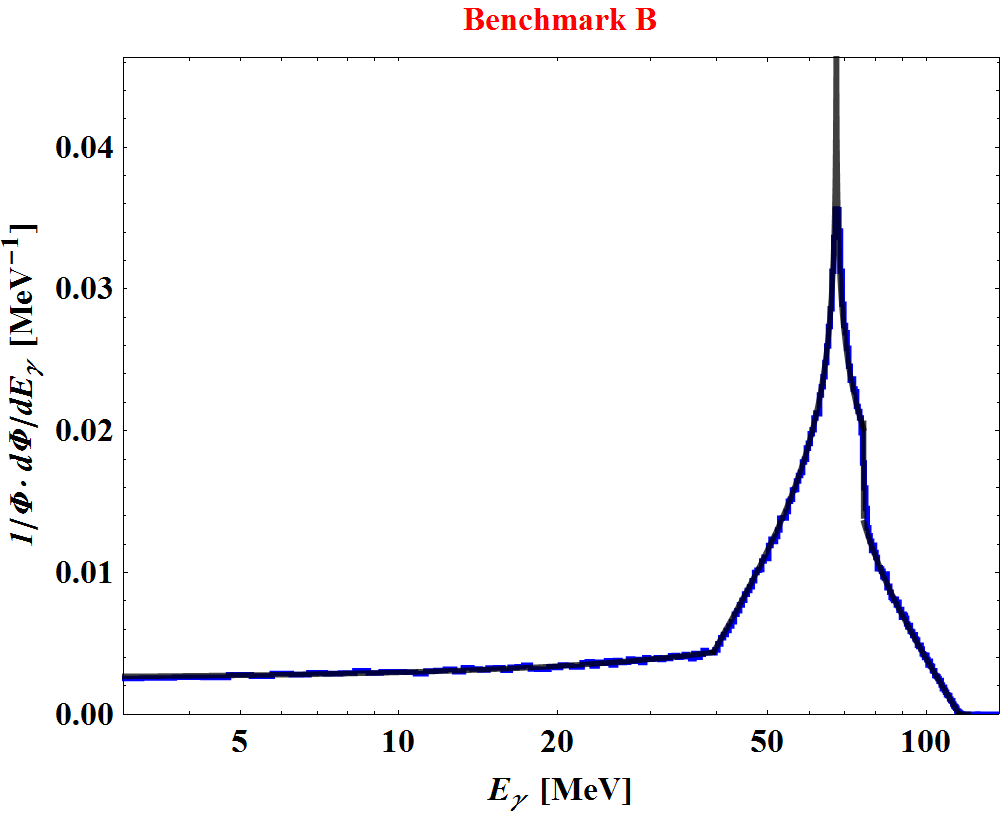}\\ \vskip 0.4cm
\includegraphics[width=0.4\textwidth, keepaspectratio]{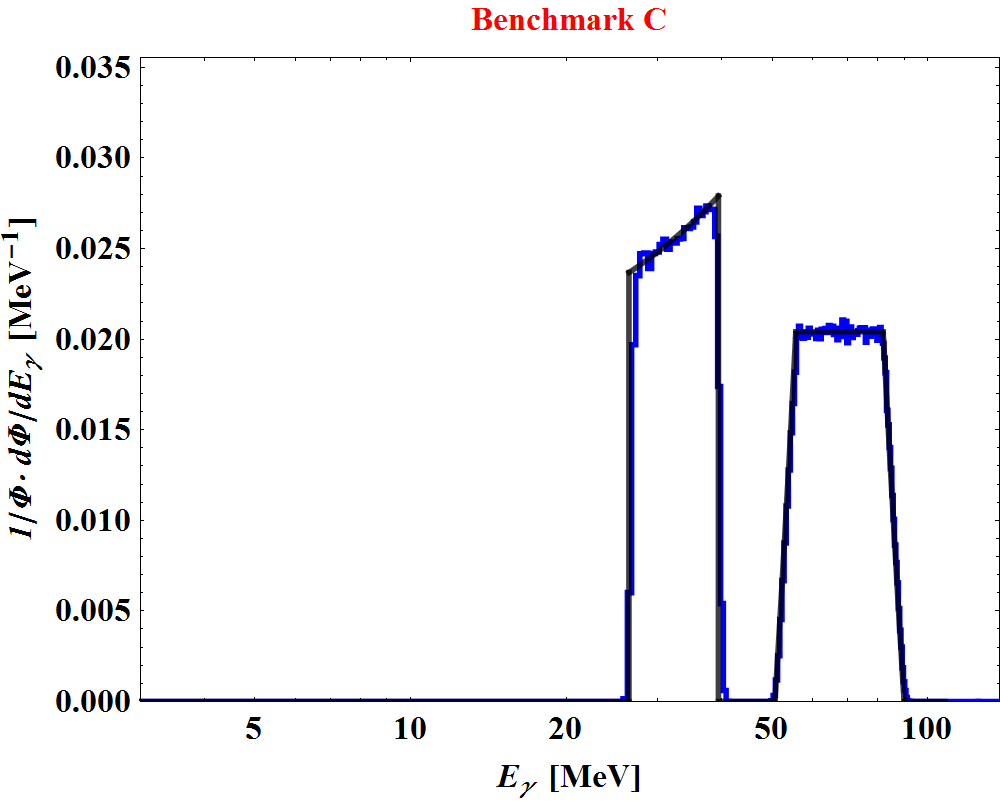}\hspace{1.2cm}
\includegraphics[width=0.4\textwidth, keepaspectratio]{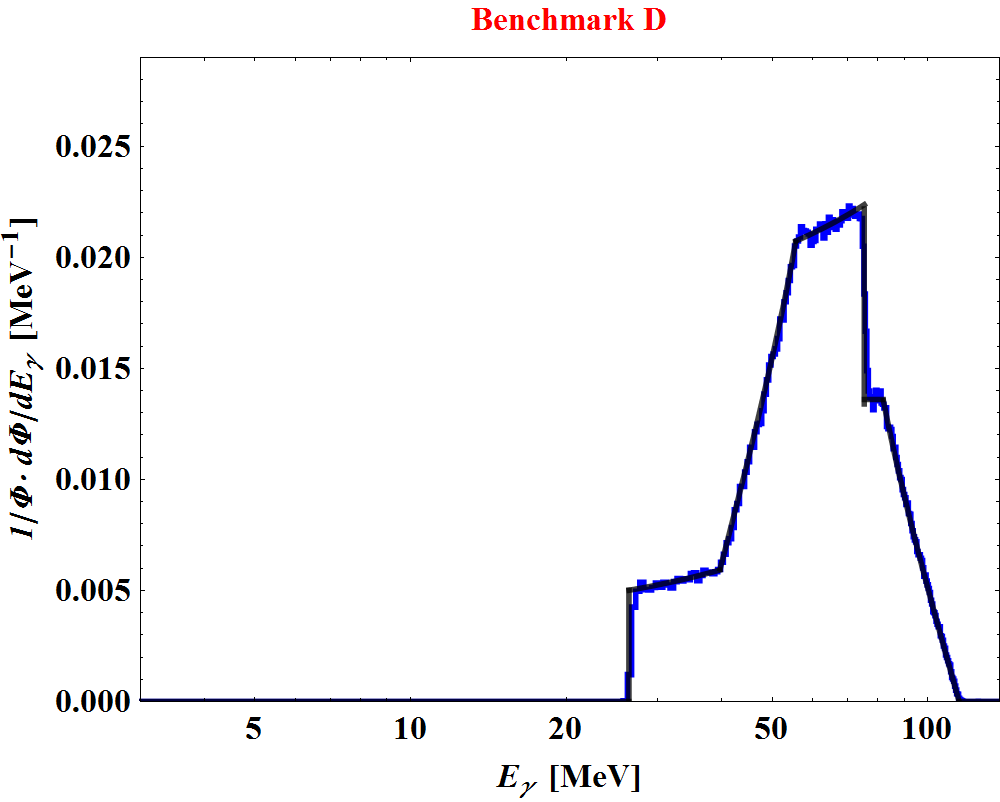}
\caption{The differential photon energy spectra associated
  with the four benchmark parameter choices $A$ through $D$ defined in Table~\protect\ref{tab:case},
   where we have taken $\xi=1$.
   The black curve in each panel
  represents the analytic result obtained by superposing the contributions to
  the photon spectrum given in Eqs.~(\ref{eq:PhiP}) and~(\ref{eq:PhiS}), while
  the blue histogram represents the results of a simulated data set smeared according
  to the Gaussian smearing function in Eq.~(\protect\ref{eq:smearing}).}
\label{fig:thcurve} 
\end{figure*}

In order to illustrate the range of different combinations of spectral shapes which
can arise in scenarios of this sort, we introduce a set of benchmark
parameter choices which exemplify four qualitatively different kinds of spectra.
The values of $\sqrt{s_0}$ and $\sqrt{s_N}$ for these benchmarks
are given in Table~\ref{tab:case}.~
For each benchmark we have taken
$\Delta \sqrt{s}=2$~MeV;
in this connection we recall that $\mp\approx 135$~MeV,
whereupon $\sqrt{2}\mp\approx 191$~MeV.~
Note that when discussing fluxes,
we shall describe our DDM ensembles in terms of the CM energies $\sqrt{s_n}$  
characterizing the annihilations/decays of their constituents 
rather than in terms of their corresponding masses $m_n$.
We do this in recognition of the fact that under
the assumption given in Eq.~(\ref{eq:flux}),
 the photon fluxes 
that result from such annihilations or decays 
depend on these CM energies rather than on the underlying masses.
In particular, by describing our ensembles in terms of CM energies rather than masses,
we retain maximal generality and need not specify whether our ensemble constituents
are annihilating or decaying.   
Indeed, this information cannot be gleaned from photon fluxes
alone, and it is only in mapping our CM energies $\sqrt{s_n}$ back to underlying masses $m_n$
that this information would be required.

The gamma-ray spectra corresponding to the benchmarks in Table~\ref{tab:case}
are displayed in Fig.~\ref{fig:thcurve}, where we have further assumed $\xi=1$. 
Note that these plots include the contributions from both primary and secondary photons.
Each of the spectra
shown in the figure has been normalized so that they all share 
a common total flux when integrated over all energies $E_\gamma$.
The black curve in each
panel represents the spectrum obtained by superposing the analytic expressions
given in Eqs.~(\ref{eq:PhiP}) and~(\ref{eq:PhiS}).  By contrast, the blue histogram
represents the results of a Monte-Carlo simulation of the corresponding
gamma-ray spectrum as they might be observed by a physical detector.  We account
for the non-zero energy resolution of the detector by smearing of the initial
photon energies obtained in the simulation using a Gaussian smearing function.
In particular, we take the probability $R_{\epsilon}$ for the detector to register
an energy $E_\gamma$, given an actual incoming photon energy $E_\gamma'$, to be
\begin{equation}
  R_{\epsilon}(E_\gamma-E'_\gamma) ~=~ \frac{1}{\sqrt{2\pi}\epsilon E'_\gamma}
  \exp\left[-\frac{(E_\gamma-E'_\gamma)^2}{2(\epsilon E'_\gamma)^2} \right]~,
  \label{eq:smearing}
\end{equation}
where $\epsilon$ is a dimensionless parameter which sets the overall scale
of the $E'_\gamma$-dependent standard deviation
$\sigma_E(E_\gamma') = \epsilon {E_\gamma'}$ of the Gaussian.  The results
in Fig.~\ref{fig:thcurve} correspond to a 1\% Gaussian smearing --- \ie, to
the choice $\epsilon =0.01$.

Benchmark~A (top left panel of Fig.~\ref{fig:thcurve}) is representative of the regime
in which $\sqrt{s_0} \approx m_{\pi^0}$ and $\sqrt{s_N} < \sqrt{2}\mp$.  In
this regime, there is no overlap between the features associated with the
contributions from primary and secondary photons, while the feature associated
with the secondary photons appears as a spike or peak rather than a plateau.
By contrast, Benchmark~B (top right panel) is representative of the regime in which
$\sqrt{s_0} \approx m_{\pi^0}$ and $\sqrt{s_N} > \sqrt{2}\mp$: 
the feature associated with the secondary photons likewise appears as a spike, but
there is a significant overlap between this feature and the feature
associated with the primary photons.  Benchmark~C (bottom left panel) is
representative of the regime in which $\sqrt{s_0}$ is significantly larger than
$m_{\pi^0}$ and $\sqrt{s_N} < \sqrt{2}\mp$:   in this regime  the features associated
with primary and secondary photons do not overlap, but the feature from
secondary photons exhibits a plateau rather than a spike.  Finally, Benchmark~D
(bottom right panel) is representative of the regime in which $\sqrt{s_0}$ is
significantly larger than $m_{\pi^0}$ and $\sqrt{s_N} > \sqrt{2} \mp$:  in this
regime the feature associated with the secondary photons likewise appears as a
plateau, but there is a significant overlap between this feature and the
feature associated with the primary photons.


\section{Discovery Reach of Future Experiments\label{sec:prospects}}


We now turn to examine the projected sensitivity of future gamma-ray experiments
to DDM ensembles which annihilate/decay primarily to $\gamma\pi^0$, followed by a subsequent
decay $\pi^0\to \gamma\gamma$.    Indeed, a
variety of proposals have recently been advanced for experiments that would significantly improve
the sensitivity to photon signals in the relevant energy range.
These include the Advanced Compton Telescope (ACT)~\cite{ACT}, the Advanced Pair
Telescope (APT)~\cite{APT}, the Gamma-Ray Imaging, Polarimetry and Spectroscopy
(GRIPS) detector~\cite{GRIPS}, the Advanced Energetic Pair Telescope (AdEPT)~\cite{AdEPT},
the Pair-Production Gamma-Ray Unit (PANGU)~\cite{PANGU}, the Compton Spectrometer
and Imager (COSI)~\cite{COSI}, and the ASTROGAM detector~\cite{ASTROGAM}.

In our analysis, for concreteness, we consider a hypothetical space-based
detector with attributes similar to those of ASTROGAM.~  In particular, we
assume that our detector is sensitive in the energy range
$0.3~\mev \lesssim E_\gamma \lesssim  3000~\mev$, and we account for the
energy resolution of the detector using a Gaussian
smearing function of the form given in Eq.~\eqref{eq:smearing}.  For simplicity,
we take the energy resolution to be 1\% (\ie, we take $\epsilon =0.01$) and we take
the effective area to be $500~\mathrm{cm^2}$ throughout this entire $E_\gamma$
range.  
These assumptions represent optimistic projections from the ASTROGAM design 
specifications, and the actual detector response will be different.
In particular, since ASTROGAM will utilize two detector technologies
in order to cover different portions of this same $E_\gamma$ range, its energy
resolution and effective area will depend non-trivially on $E_\gamma$.

Our goal is to assess the discovery reach of our hypothetical detector 
as a function of the parameters governing our underlying DDM model.
We shall assess this discovery reach as follows. 
First, we define the quantity
\beq
         \frac{\tilde \Phi}{J} ~=~  \frac{4\pi}{3J} \sum_n \Phi_n ~=~ 
          \sum_n \frac{\Omega_n}{\Omega_{\rm tot}}
                 \frac{\lambda_n}{m_n} ~\equiv ~ \langle \lambda/m \rangle
\eeq
where $\tilde \Phi \equiv  (4\pi/3)\Phi$ is the normalized total flux that we would
expect to see from a given DDM model. 
In some sense
this quantity represents the ``particle-physics'' contribution to the total flux, with the
astrophysical factor $J$ divided out.
In order to assess the reach of our hypothetical detector, we therefore seek the
critical (minimal) value of $\tilde \Phi/J$ 
for which an excess  might become apparent
after one year of continuous observation.
Or, phrased conversely, we 
seek to determine the maximum value of $\tilde\Phi/J$
for which {\it no}\/ appreciable signal can be resolved after one year of continuous
observation.  
If this maximum value of $\tilde \Phi/J$ is relatively small for a given 
set of underlying DDM parameters, 
our telescope is extremely sensitive to the corresponding DDM photon flux 
and our discovery reach is enhanced.
By contrast, if this maximum value of $\tilde \Phi/J$ is relatively large,
the corresponding discovery reach of our hypothetical telescope is suppressed.

In our analysis, we shall consider two different regions of interest
on the sky 
which correspond to two of the most promising search strategies for gamma-ray
signals of dark-matter annihilation/decay: searches in dwarf spheroidal galaxies
and searches in the diffuse galactic gamma-ray spectrum.  We do not consider
signals from the Galactic Center, as the astrophysical backgrounds in this
region are not well understood and systematic uncertainties are therefore
expected to be large.

\subsection{Dwarf-spheroidal search\label{sec:DwarfSearch}}

Dwarf spheroidal galaxies provide a particularly auspicious environment
in which to search for signals of annihilating/decaying dark matter.
Observations of stellar kinematics within these galaxies suggest that they
are highly dark-matter dominated~\cite{MateoDwarfs,McConnachieDwarfs}.
In addition, since the solid angle on the
sky subtended by many of these galaxies is small, reasonably reliable empirical
estimates of the astrophysical foregrounds and backgrounds can be obtained from
measurements of the differential gamma-ray flux in the surrounding region.
Moreover, since most known dwarf spheroidals lie at significant distances from
the galactic plane of the Milky Way, these astrophysical foregrounds are small.

For concreteness, we focus our analysis on one particular dwarf galaxy,
Draco, which subtends a solid angle of approximately $1.6\times 10^{-3}$~sr
on the sky.  For a region of interest defined by this solid angle,
an empirical reconstruction of the dark-matter halo profile of
this galaxy from stellar-kinematic data~\cite{GeringerSameth:2014yza} yields
a $J$-factor
$\log_{10} (J / \mathrm{GeV}^2\mathrm{cm}^{-5}) = 19.05^{+0.22}_{-0.21}$
for annihilation and
$\log_{10} (J / \mathrm{GeV}\mathrm{cm}^{-2} ) = 18.97^{+0.17}_{-0.24}$
for decay.  For simplicity, we assume that the main source of
foreground/background photons is diffuse emission and assume that
contributions from nearby extragalactic sources are negligible.
We model the diffuse
contribution to the differential gamma-ray flux using a single power law,
with a normalization coefficient and scaling index derived from a
fit to COMPTEL~\cite{Weidenspointner:1999thesis} and
EGRET~\cite{Strong:2004ry} data:
\begin{equation}
  \frac{d^2\Phi_b}{dE\, d\Omega} ~=~ 2.74 \times 10^{-3}
  \left(\frac{E}{\mathrm{MeV}}\right)^{-2.0}
  \mathrm{cm}^{-2}  \mathrm{s}^{-1}  \mathrm{sr}^{-1} \mathrm{MeV}^{-1}.
  \label{eq:diffuse-bkg}
\end{equation}

In general, the DDM discovery reach of our hypothetical detector 
depends on the underlying DDM parameters $\sqrt{s_0}$, $\sqrt{s_N}$, and $\xi$.
(As usual, we are assuming $\delta=1$ and $\Delta \sqrt{s} = 2$~MeV.)~
For each choice of parameters, 
our results in Eqs.~(\ref{eq:PhiP}) and~(\ref{eq:PhiS})
make a prediction concerning the signal differential fluxes 
$d\Phi^{(s,p)}/dE_\gamma$
of primary and secondary photons,
respectively.  In particular, for any given values of 
$(\sqts0,\sqrt{s_N})$, these signal fluxes 
stretch over only a finite range of energies $E_\gamma$.
Thus, for any given $(\sqts0,\sqrt{s_N})$, we shall restrict our 
analysis to those energy bins lying within this range.

The choice of 
$(\sqts0,\sqrt{s_N},\xi)$ determines the overall {\it shape}\/
of the signal differential flux as a function of photon energy $E_\gamma$,
while the overall magnitude of this differential flux is determined by the 
normalization $\Phi_0$.
Thus, for any given choice of 
$(\sqts0,\sqrt{s_N},\xi)$, 
we then seek to find the critical (minimal) value of $\Phi_0$
for which an excess signal just becomes observable.
Equivalently, we seek the largest value of $\Phi_0$ for which {\it no}\/ signal can be
discerned.
This largest value of $\Phi_0$ then leads to a corresponding largest value of $\tilde\Phi/J$, 
where the numerical value of $J$ is given above.

In general, there are two different paths we might follow in order to determine
this critical value of $\Phi_0$.
One possible procedure is to find the critical value of
$\Phi_0$ 
for which an excess {\it in any single bin}\/ just becomes observable
(or equivalently, the largest value of $\Phi_0$ for which no signal can be
discerned {\it in any single bin}\/).
Within each bin, observability would be assessed as follows.
In general, the expected number of events within a given bin
includes a signal contribution from DDM annihilation/decay
within the halo of Draco as well as a background contribution given
by Eq.~\eqref{eq:diffuse-bkg}.  
We would then seek the maximum value of $\Phi_0$ for which this observed
number of events in every bin is consistent 
with the contribution from background alone within 95\%~C.L., assuming
Poisson statistics.

The above procedure describes a ``binned'' approach to determining the critical value 
of $\tilde \Phi/J$ which is sensitive
to the overall {\it shape}\/ of the differential flux --- \ie, 
an approach which is based on an analysis of the 
counts within individual energy bins.
However, an alternative path is to simply focus instead on the total integrated flux across all energy bins, and 
to determine the critical value of $\tilde \Phi/J$ for which 
this integrated flux exceeds the integrated contribution from background
alone within 95\%~C.L., assuming
Poisson statistics.

In order to asssess the greatest (maximum) discovery reach, we shall employ
whichever method (binned or integrated)
yields the smallest value for $\tilde\Phi/J$. 
It turns out that if
$\sqrt{s_0} \approx m_{\pi^0}$, the primary photon spectrum extends down to very low photon energies where the diffuse background is quite large. Incorporating these high-background bins into a total (integrated) counting analysis significantly weakens the estimate of the discovery reach.  Consequently, for $\sqrt{s_0}\approx \mp$, it turns out that the 
binned analysis yields a greater discovery reach.  For larger values of $\sqrt{s_0}$,
by contrast, it turns out that an analysis based on the total integrated flux is superior.

\subsection{Diffuse-background search\label{sec:DiffuseSearch}}

The total diffuse gamma-ray background consists of a contribution from unresolved
astrophysical sources as well as both a galactic and an extragalactic contribution
from dark-matter annihilation/decay.  The extragalactic dark-matter contribution
is assumed to be isotropic, while the galactic contribution depends (through the
$J$-factor) on the dark-matter halo profile of the Milky Way.
However, this latter contribution is not particularly
sensitive to the form of the inner halo profile in situations in which the region of interest
includes only areas of the sky far from the Galactic Center.  Moreover, the
diffuse extragalactic contribution to the photon flux from any particular
location on the sky is typically subleading in comparison with the diffuse
galactic contribution, except for cases in which that location lies near either of
the galactic poles (where the latter contribution is presumably at its minimum).
Accordingly, we adopt as our region of interest the region in which the galactic
latitude $b$ lies within the range $20^\circ < |b| < 60^\circ$.
In the following, we calculate the $J$-factors from their differential forms for an NFW profile, 
for which numerical evaluations are given in Ref.~\cite{Jfactor882360}. 

Disentangling the dark-matter contribution to the diffuse gamma-ray flux
from the astrophysical background requires detailed knowledge of that
background.  However, the astrophysical contribution to the diffuse gamma-ray
flux is not well measured or understood.  Given this uncertainty,
we evaluate the discovery reach for this diffuse search using two different
methods.  The first of these involves no assumptions about the astrophysical
background and yields a more conservative estimate of the discovery reach,
while the second assumes a particular functional form for the background
and thereby yields a more optimistic estimate.

In deriving our more conservative estimate of the discovery reach, we compare the gamma-ray flux
spectrum observed by our hypothetical detector to the expected signal contribution
from dark-matter annihilation/decay alone.  More specifically, we compare the
number of events observed in each energy bin to the corresponding number of
expected events, given a particular choice of DDM model parameters.  Under the
assumption that the observed number of events in each bin is given by the
background spectrum in Eq.~\eqref{eq:diffuse-bkg}, we derive an upper limit
on $\tilde \Phi/J$ for which this observed number of events
$\mathcal{N}_i^{\mathrm{obs}}$ in each bin is consistent with the theoretical expectation
$\mathcal{N}_i^{\mathrm{exp}}$ to within $2\sigma_i$, where the index $i$ labels the bin
and where $\sigma_i$ denotes the corresponding uncertainty.  In particular, $\sigma_i$
is dominated by systematic uncertainty in the expression for the differential
flux in Eq.~\eqref{eq:diffuse-bkg}, which we take to be 15\% of the flux itself.

In deriving our more optimistic estimate of the discovery reach, we follow
a procedure which is similar to that followed for the dwarf-spheroidal search.  However, 
rather than neglecting the background contribution to the expected
number of events in each bin, in this case we assume that this background contribution is
given by Eq.~\eqref{eq:diffuse-bkg}.  Once again, we derive an upper limit
on $\tilde \Phi/J$ by assuming that the observed
number of events in each bin is likewise given by the background spectrum
in Eq.~\eqref{eq:diffuse-bkg} and requiring consistency between
$\mathcal{N}_i^{\mathrm{obs}}$ and $\mathcal{N}_i^{\mathrm{exp}}$ to within $2\sigma_i$ in each bin.

\subsection{Results}

\begin{figure*}
  \centering
  \includegraphics[width=0.95\textwidth, keepaspectratio]{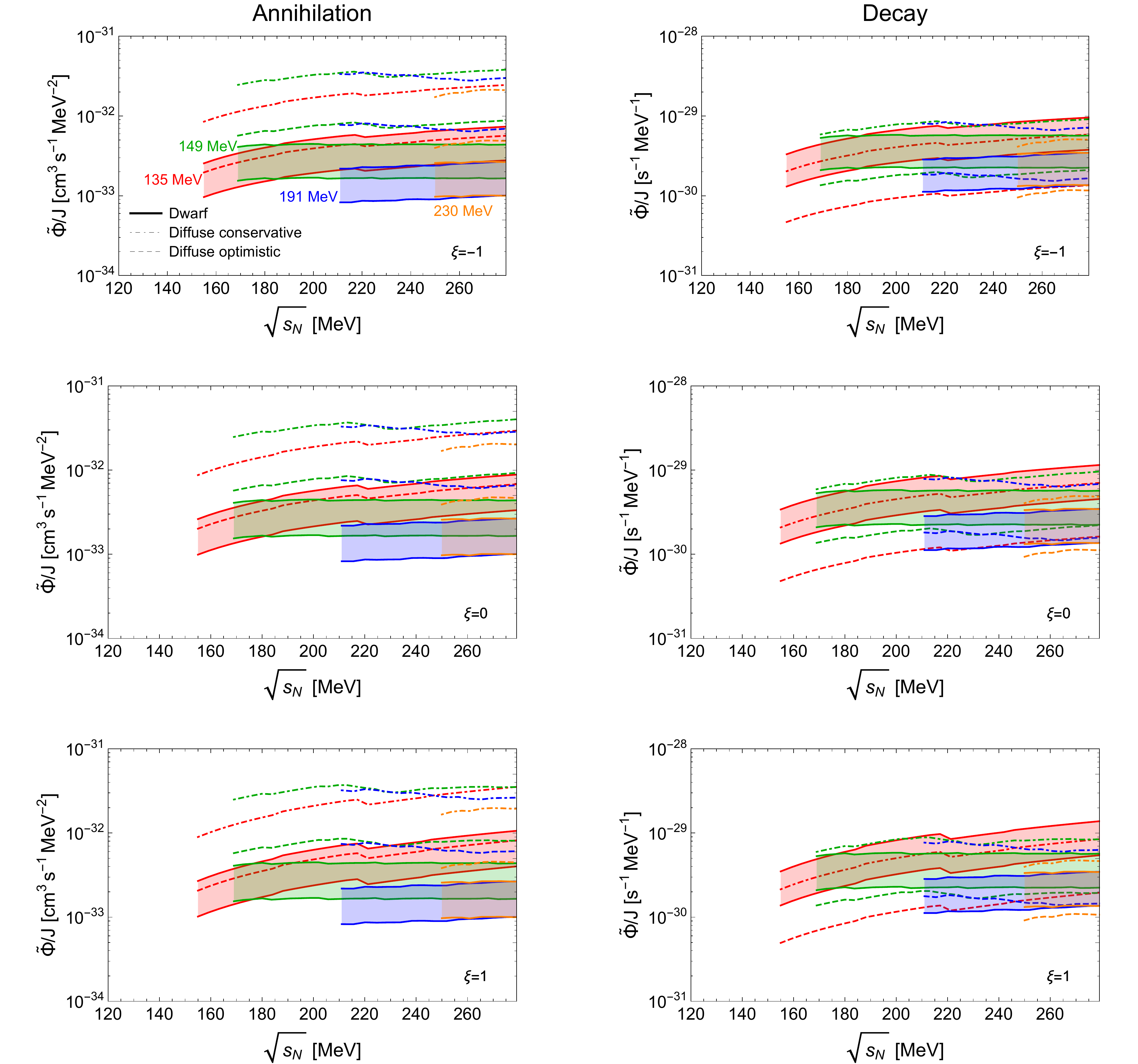}
  \caption{The projected discovery reach for a representative next-generation
    MeV-range gamma-ray telescope, plotted as functions of $\sqrt{s_N}$ for
    different values of $\sqrt{s_0}$ and $\xi$, with $\Delta(\sqrt{s}) = 2$~MeV.~  
    The results are shown as an upper limit on the quantity $\tilde \Phi/J$ for
    which a statistically significant signal is not observed within one year of
    continuous observation.  
    Panels in the top, middle, and bottom rows correspond to $\xi= -1,0,+1$, respectively,
    while those in the left and right columns correspond respectively to annihilating and decaying
    dark-matter scenarios.
    Within each panel, four benchmark choices of $\sqrt{s_0}$ are shown: $\sqrt{s_0}=135~\mev$ (red curves),
    $149~\mev$ (green curves), $\sqrt{s_0}=191~\mev$ (blue curves), and
    $\sqrt{s_0}= 230~\mev$ (orange curves).  
    In each case we then show results for $\sqrt{s_N}$
     within the range $\sqrt{s_0}+ 10\Delta\sqrt{s} \leq \sqrt{s_N} \leq 2 m_{\pi^\pm}$.
     The solid bands shown in each panel
    correspond to the results of the dwarf-spheroidal search, as outlined in
    Sect.~\protect\ref{sec:DwarfSearch}, 
      with the results for $\sqrt{s_0}=135$~MeV obtained
      through a binned approach and the others obtained through 
      an approach based on the total integrated flux.
    The width of each band reflects a
    $1\sigma$ uncertainty in the $J$-factor for the dwarf.  The dashed and
    dot-dashed lines correspond to the results of a diffuse-background search
    using the optimistic and conservative analysis methods outlined in
    Sect.~\protect\ref{sec:DiffuseSearch}, respectively.
    \label{fig:exclusion-sqrtSN}}
\end{figure*}

The discovery reaches for both the dwarf-spheroidal search and the diffuse-background
search are shown in Fig.~\ref{fig:exclusion-sqrtSN}.  In this
figure, the bounds on $\tilde \Phi/J$ from each search are shown as a
function of the parameter $\sqrt{s_N}$ for the four different reference values of
$\sqrt{s_0}$ labelled within each panel, with 
$\Delta\sqrt{s} = 2$~MeV
and $\sqrt{s_0}+ 10\Delta\sqrt{s} \leq \sqrt{s_N} \leq 2 m_{\pi^\pm}$.
This lower bound on $\sqrt{s_N}$ ensures that we are including the contributions
of at least 10 ensemble constituents $\phi_n$
in addition to $\phi_0$ for each chosen value of $\sqrt{s_0}$, while
the upper bound  
ensures that we do not exceed the threshold
$2m_{\pi^\pm}$ for charged-pion pair-production, beyond which
additional flux contributions must be included.
Results for $\xi= -1, 0, +1$ are shown along the top, middle, and bottom rows of Fig.~\ref{fig:exclusion-sqrtSN},
while the panels within the left and right columns of Fig.~\ref{fig:exclusion-sqrtSN} show the results for
annihilating and decaying dark-matter scenarios respectively.  
The solid colored bands indicate the results of the dwarf-spheroidal search, with
the width of each band
reflecting a $1\sigma$ uncertainty in the $J$-factor for the dwarf.  By contrast,
the dashed and dot-dashed lines correspond to the results of a diffuse-background
search using the optimistic and conservative analysis methods outlined in
Sect.~\protect\ref{sec:DiffuseSearch}, respectively.

For the dwarf-spheroidal search, 
the results shown in Fig.~\ref{fig:exclusion-sqrtSN} indicate that the
discovery reach for our hypothetical telescope 
tends to be relatively insensitive to $\sqrt{s_N}$ for large $\sqrt{s_0}$,
but more sensitive to $\sqrt{s_N}$ for smaller $\sqrt{s_0}$.
When scanned over possible values of $\sqrt{s_0}$, however,
the discovery reach tends to be relatively insensitive to $\sqrt{s_N}$: 
cases with large $\sqrt{s_0}$
provide the greatest reach 
when $\sqrt{s_N}$ is large but 
cases with smaller $\sqrt{s_0}$ provide the greatest reach when $\sqrt{s_N}$ is smaller. 

It is also noteworthy that when $\sqrt{s_0}\approx \mp$,
it is the binned analysis which provides the greater discovery reach;
the opposite is true when $\sqrt{s_0}$ is larger.
However, this can be understood as follows.
When $\sqrt{s_0} \approx m_{\pi^0}$, the primary photon spectrum extends down to 
very low gamma-ray energies where the diffuse background is quite large. 
Incorporating these high-background bins into an analysis based on total counts 
then significantly weakens the estimate of the discovery reach.
However, this feature does not affect the individually-binned analysis where the 
effects from such low-energy bins no longer dominate the analysis.
Thus, in such cases, the results of the binned analysis 
are stronger than those of an integrated analysis.
Indeed, this is ultimately why the overall discovery reach
for small $(\sqrt{s_0},\sqrt{s_N})$ remains competitive 
with that for larger values 
of $(\sqrt{s_0},\sqrt{s_N})$, as shown in Fig.~\ref{fig:exclusion-sqrtSN}.
Indeed, we see from Fig.~\ref{fig:exclusion-sqrtSN}
that this remains true for all of the values of $\xi$ surveyed.

For the diffuse-background search, by contrast,
the discovery reach depends
more stikingly on both $\sqts0$ and $\sqrt{s_N}$.  
Moreover, the reach is sensitive to
the spectral shapes of both the primary  and secondary photon contributions to
the gamma-ray spectrum.  Overall, the secondary photon contribution has a
more significant impact on the discovery potential.  The reason is that
in the regime in which 
$\sqrt{s_N}$ is reasonably small and
$\sqrt{s_0} \approx m_{\pi^0}$,
the secondary photon spectrum is sharply peaked around
$E_\gamma = m_{\pi^0}/2$.  As a result, the potential for observing an excess
in the corresponding energy bin has a profound positive effect on the overall
discovery reach.  Indeed, it is evident from Fig.~\ref{fig:exclusion-sqrtSN}
that the reach is greatest in the regime in which 
$\sqrt{s_N}$ is relatively small and
$\sqts0 \approx m_{\pi^0}$.
As $\sqrt{s_0}$ increases away from $m_{\pi^0}$ and the peak
becomes a plateau, the potential for observing an excess in this bin decreases.
Increasing $\sqrt{s_N}$ for fixed $\sqrt{s_0}$ has the effect of broadening
the secondary photon spectrum.  On the one hand, this broadening reduces the
significance of the peak at $E_\gamma = m_{\pi^0}/2$; on the other hand,
it also extends the upper edge of the secondary photon spectrum to higher
$E_\gamma$, where the astrophysical background is smaller and a signal is
more readily observable.  As a result of the interplay between these two
effects, the discovery reach initially falls with increasing $\sqrt{s_N}$ because the
energy bin corresponding to the peak provides the best prospects for observing
an excess in signal events.  However, as $\sqrt{s_N}$ increases further and the
higher-energy bins become the most relevant for observing an excess, the
discovery potential stabilizes.

While the role played by the primary photon spectrum 
in determining the discovery reach for the diffuse-background search 
is less pronounced than that played 
by the secondary photon spectrum, 
the primary photon spectrum still has a demonstrable effect on the
discovery reach.  In particular, as $\sqts0$ increases, the primary photon
spectrum is shifted to higher values of $E_\gamma$ where astrophysical
backgrounds are small.  For sufficiently large $\sqts0$, this effect
more than compensates for the corresponding broadening of the secondary photon
spectrum and yields an overall increase in the discovery reach.

Comparing the cases of dark-matter annihilation and decay, we see that
the dwarf search has an order-of-magnitude greater discovery reach than 
the diffuse search for annihilation, while both searches have comparable 
discovery reaches for decay. Since the $J$-factor in Eq.~(\ref{Jdef}) 
depends on $\rho^2$ for annihilation, we expect the dense environment of the dwarf to be a more 
advantageous system in which to search for annihilating dark matter 
than the diffuse background. For decay, however, the $J$-factor involves only 
a single power of $\rho$, and thus the dwarf search does not possess the 
same upper hand as it has for annihilation.


\section{Extraction of Dark-Sector Parameters\label{sec:measurement}}


As discussed in the Introduction, our primary motivation for studying
DDM ensembles whose constituents annihilate/decay primarily into a $\gamma\pi^0$
final state, followed by the subsequent decay $\pi^0\to \gamma\gamma$, 
is that the shapes of the spectral features associated with
primary and secondary photons are correlated.   A comparison between the information
extracted from these two features can therefore provide a powerful consistency check
on the DDM interpretation of such a gamma-ray excess.  
However, we are not merely 
interested in the prospects for {\it observing}\/ a signal of a DDM ensemble with this
annihilation/decay phenomenology, as in Sect.~\ref{sec:prospects};  
we are also interested in determining the degree to which we might then 
 {\it extract}\/ the values of the underlying parameters which characterize the DDM ensemble. 
This is the subject to which we now turn.

Towards this end, we shall focus on the four benchmarks outlined in 
Table~\ref{tab:case} and illustrated in Fig.~\ref{fig:thcurve} with $\xi = 1$.
For each benchmark, we shall  
investigate the prospects for extracting the corresponding underlying DDM model parameters
$(\sqrt{s_0},\sqrt{s_N},\xi)$ 
by generating and then analyzing corresponding sets of simulated detector data.
We begin our discussion by outlining how these data sets are 
generated and analyzed.
We then discuss the extent
to which our underlying DDM parameters can be meaningfully extracted in each case.
Specifically, using the simulated detector data for each benchmark, we shall focus
on two critical but somewhat distinct questions:
\begin{itemize}
\item    To what extent can we extract {\it evidence}\/ of a correlation between 
               primary and secondary photon flux spectra?
\item    To what extent does the {\it assumption}\/ of such a correlation 
               {\it enhance}\/ our ability to extract the corresponding 
               underlying DDM model parameters?
\end{itemize}
Note that a positive outcome to the first question 
implicitly strengthens our interpretation of a measured photon flux  
as resulting from annihilating/decaying dark matter (as opposed to, say, 
other astrophysical sources).
By contrast, once we are assured that such a photon flux has a dark-matter origin,
such a correlation between the primary and secondary photon fluxes is {\it automatic}\/.   
It is then the second question above which becomes critical for extracting the underlying
physics of the dark sector.

\subsection{Generating and analyzing simulated data sets}

In order to generate our simulated data sets,
we begin by determining the total expected number $N_B$ of background events observed
by our hypothetical detector within our region of interest during one year
of continuous observation.  
This number $N_B$ is therefore evaluated across the entire energy range $0.3~\mev < E_\gamma < 3000~\mev$  
to which the detector is sensitive, 
yielding the result $N_B\approx 2.32\times 10^5$.
Likewise, we determine a number $N_S$ of signal events by assuming the
minimum necessary in order to claim a $5\sigma$ discovery based on a simple counting
analysis in which the statistical significance is estimated using $N_S/\sqrt{N_B}$.
This yields $N_S\approx 5 \sqrt{N_B} \approx 2.41\times 10^3$.
In principle, one might argue that the values of $N_B$ and $N_S$ should depend
on the energy range over which the particular benchmark can be expected to provide data and thereby
be sensitive to background.
However, since there are relatively few background events in the high-energy regime,
it turns out that the above values of $N_B$ and $N_S$,
as calculated for our hypothetical detector as a whole,
are not significantly different from those that would correspond to Benchmark~B, which has the largest energy range.
In the following, we will take the above values of $N_B$ and $N_S$ to be fixed across all benchmarks.  
This allows us to make
a meaningful comparison across benchmarks by considering our fixed quantity to be the
number of signal events itself (rather than, say, a corresponding statistical significance).
This procedure for calculating signals and backgrounds across the entire energy range 
to which our hypothetical detector is sensitive also reflects
what one would actually do upon faced with an experimental signal ---
namely, analyze this signal over the entire energy range available, without any
foreknowledge or assumptions regarding the particular underlying spectral features involved.

Given the above values of $N_B$ and $N_S$,
the generation of our simulated data set for each benchmark proceeds as follows.
The signal contribution associated with each ensemble constituent is determined by
partitioning the $N_S$ signal events among the $\phi_n$ in proportion to the contribution $\Phi_n$
that each makes toward the total photon flux $\Phi$.  Photon energies for background
events are generated randomly from the relevant probability distribution function 
over the entire range mentioned above.
Photon energies for the set of signal events associated with a given $\phi_n$ are
also generated randomly, with one third of the events assigned
the primary photon energy $E_{\rm line}$ given in Eq.~(\ref{Eline}) and the
other two thirds distributed according to a normalized probability distribution
function derived from Eq.~(\ref{twothetas}).
Finally, the raw $E_\gamma$ values for both signal and background events are
smeared according to Eq.~(\ref{eq:smearing}) with $\epsilon = 0.01$ in
order to account for the energy resolution of the detector.

The net result of this procedure is a set of four simulated energy spectra
that might emerge from the decays/annihilations of our four DDM ``benchmark'' ensembles.
Our analysis of these data sets then proceeds as follows.
First, recognizing that these data sets represent the total ``observed''
differential photon fluxes,
we begin by disentangling our ``signal'' contribution 
from astrophysical backgrounds.  For this reason, we focus exclusively on
dwarf-spheroidal searches, as the corresponding backgrounds can be estimated directly 
from measurements.  For concreteness, we consider the same region of interest
which characterized the dwarf-spheroidal search in Sect.~\ref{sec:DwarfSearch}
and adopt the same set of parameters for our hypothetical detector.  To
isolate the signal contribution, we employ a minimal background-subtraction
procedure in which an expected number of background events
${\mathcal{N}}_i^{\mathrm{BG}}$ in each energy bin is derived using
the background model in Eq.~\eqref{eq:diffuse-bkg}
and is subtacted from the corresponding total number of observed events 
${\mathcal{N}}_i^{\mathrm{Data}}$.
Again, we emphasize that we can follow this procedure because experimentalists will actually be able to measure 
the background, unlike the situation in the case of a diffuse search.
The resulting number of events 
\begin{equation}
  \mathcal{N}_i^{\mathrm{Sig}} ~\equiv ~ \mathcal{N}_i^{\mathrm{Data}}
    - {\mathcal{N}}_i^{\mathrm{BG}}~
\end{equation}
is thus our ``signal'' contribution, to be interpreted as 
coming from the decays/annihilations of the constituents of the DDM ensemble.

Given this signal contribution, we determine the corresponding values
of the underlying DDM shape parameters 
$(\sqrt{s_0}, \sqrt{s_N}, \xi)$ by fitting the template functions
in Eqs.~(\ref{eq:PhiP}) and~(\ref{eq:PhiS}) to this residual spectrum.
However, the specific fit we perform will depend on which of the fundamental
questions itemized above we are attempting to answer.

To address the first question, we perform {\it independent}\/ fits 
of the primary and secondary flux spectra, extracting independent best-fit
values $(\sqrt{s_{0,p}},\sqrt{s_{N,p}},\xi_p)$ for the primary flux spectra
and $(\sqrt{s_{0,s}},\sqrt{s_{N,s}},\xi_s)$ for the secondary flux spectra.
Comparing these sets of parameters with each other thus provides a test of our
purported correlations between these two spectra.
Likewise, comparing each independent set of parameters against our corresponding 
original benchmark values provides a measure of our ability 
to extract our underlying DDM parameters {\it without}\/ assuming a correlation between the two spectra.
By contrast, to address the second question,
we perform a constrained fit of {\it both}\/ spectra simultaneously
with only a single set of free parameters
$(\sqrt{s_0},\sqrt{s_N},\xi)$.
Comparing the results thus obtained with those previously obtained with
independent fits for each spectrum then provides a measure
of the extent to which the existence of a correlation between
the two spectra enhances our ability to extract the underlying
DDM model parameters.

In practice, 
it is important to recognize that there is actually another variable
beyond the shape variables $(\sqrt{s_0},\sqrt{s_N},\xi)$
which must also be fit when extracting our 
underlying DDM parameters:   this is the overall normalization 
factor $\Phi_0$.
In fact, strictly speaking, the overall normalization factor for both the primary photon
spectrum in Eq.~(\ref{eq:PhiP}) and
the secondary photon spectrum 
in Eq.~(\ref{eq:PhiS}) 
is not $\Phi_0$ alone, but rather the
parameter combination 
\beq
     \Psi ~\equiv~ \Xi \, (\sqrt{s_0})^{-\xi}~,
\label{Psidef}
\eeq
where
\beq
     \Xi ~\equiv~ \frac{\Phi_0}{\Delta(\sqrt{s})}~.
\label{Xidef}
\eeq
We shall therefore fit the
aggregate quantity $\Psi$ directly, and only subsequently extract a value for $\Xi$ 
using the results of our overall fits for $\sqrt{s_0}$ and $\xi$.  
Unfortunately, without {\it a priori}\/ knowledge of $\Delta(\sqrt{s})$, we see that
the parameter combination $\Xi$ cannot be disentangled further and thus 
represents the irreducible limit of our ability to extract 
the underlying DDM flux normalization using these methods.

As briefly discussed in Sect.~\ref{sec:case-studies},
the procedure that we shall use in performing these parametric fits to the signal
spectrum depends on the degree of overlap between the spectral features
associated with primary and secondary photons.  In the regime in which
$\sqrt{s_N} < \sqrt{2} \mp$, these two features are well separated
and a fit can be performed for each feature independently.
Indeed, this will be our procedure for Benchmarks~A and C.~ 
By contrast, in the regime in which $\sqrt{s_N} >  \sqrt{2} \mp$,
the overlap is significant
and a single fit must be performed for both features simultaneously.
This will be our procedure for Benchmarks~B and D.  

Thus, summarizing, the specific types of fits we shall perform 
depend not only on which of the questions itemized above
we are seeking to address, but also on which benchmark we  
are studying.
To address the first question for Benchmark~A,
we shall perform two independent four-parameter fits, extracting
independent values 
$(\sqrt{s_{0,p}},\sqrt{s_{N,p}},\xi_p,\Psi_p)$ 
and 
$(\sqrt{s_{0,s}},\sqrt{s_{N,s}},\xi_s,\Psi_s)$ 
using our data sets for 
the primary and secondary spectra respectively.
We also follow an identical procedure in order to address the first question
for Benchmark~C.~
Indeed, it is only because these two spectra are non-overlapping for Benchmarks~A and C
that we allow each fit to have its own independent normalization in these cases.
By contrast, in order to address the first question for Benchmark~B
or Benchmark~D, we perform a single seven-parameter fit to the parameters
$(\sqrt{s_{0,p}},\sqrt{s_{N,p}},\xi_p,
  \sqrt{s_{0,s}},\sqrt{s_{N,s}},\xi_s,\Psi)$.
Indeed, in these cases, the overlapping nature of the primary  and secondary photon 
spectra requires that we impose a common normalization $\Psi$ during the fitting
process.
Of course, the results of this fit then yield independent values for
$\Xi_{p} = \Psi (\sqrt{s_{0,p}})^{\xi_p}$
and 
$\Xi_{s} = \Psi (\sqrt{s_{0,s}})^{\xi_s}$. 
Finally, in order to address the second question for each benchmark, we compare
the above results with those obtained through a single four-parameter fit
to the underlying DDM parameters
$(\sqrt{s_0},\sqrt{s_N},\xi,\Psi)$.

Note that this analysis applies equally well for either annihilation or decay, 
as the only difference between
these two cases lies not in the extracted values of the $\sqrt{s_n}$ parameters but 
rather in the subsequent mapping between these parameters and the 
original DDM mass variables $m_n$,
as already discussed in Sects.~\ref{sec:model} 
and \ref{sec:spectrum} [especially Eq.~(\ref{sndef})]
and at the end of the Appendix.


\subsection{Results}

\begin{figure*}[t]
\includegraphics[width=0.425\textwidth, keepaspectratio]{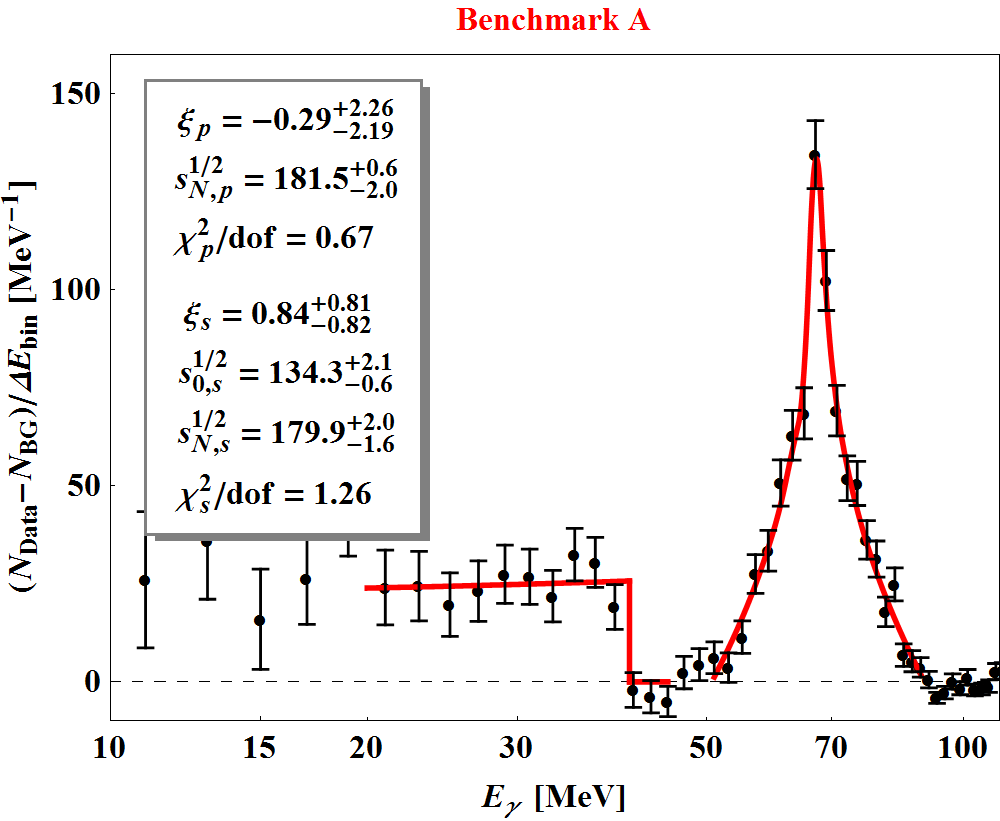} \hskip 1.0cm
\includegraphics[width=0.425\textwidth, keepaspectratio]{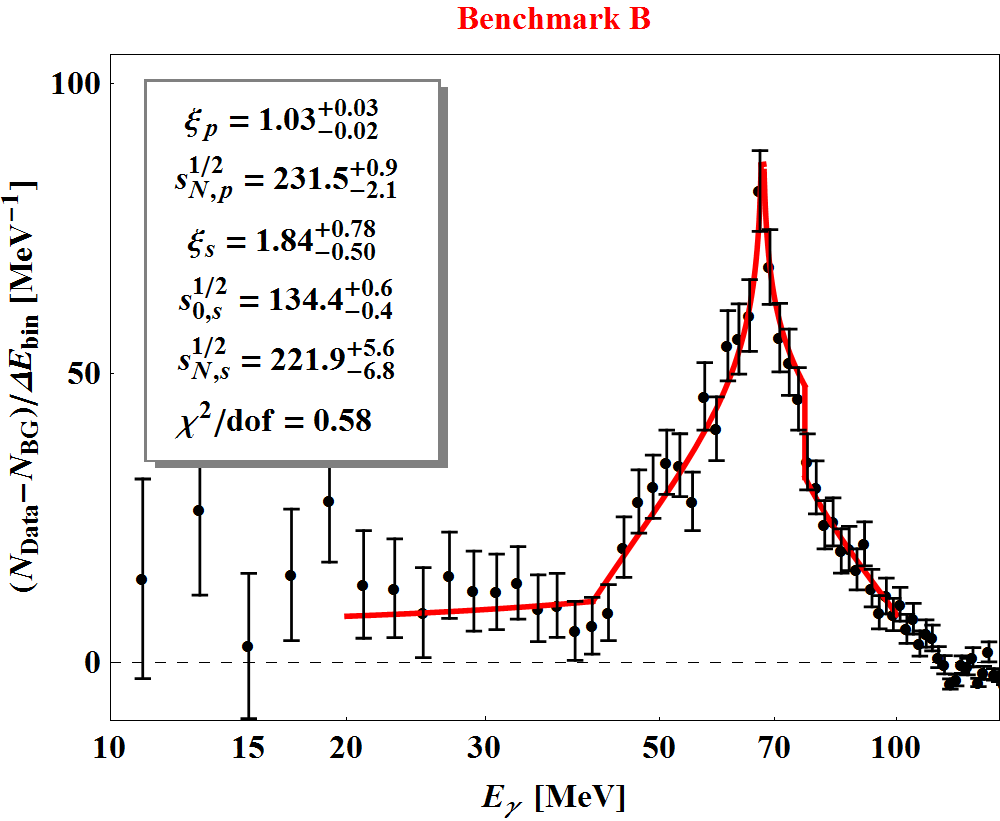} \vskip 0.4cm
\includegraphics[width=0.425\textwidth, keepaspectratio]{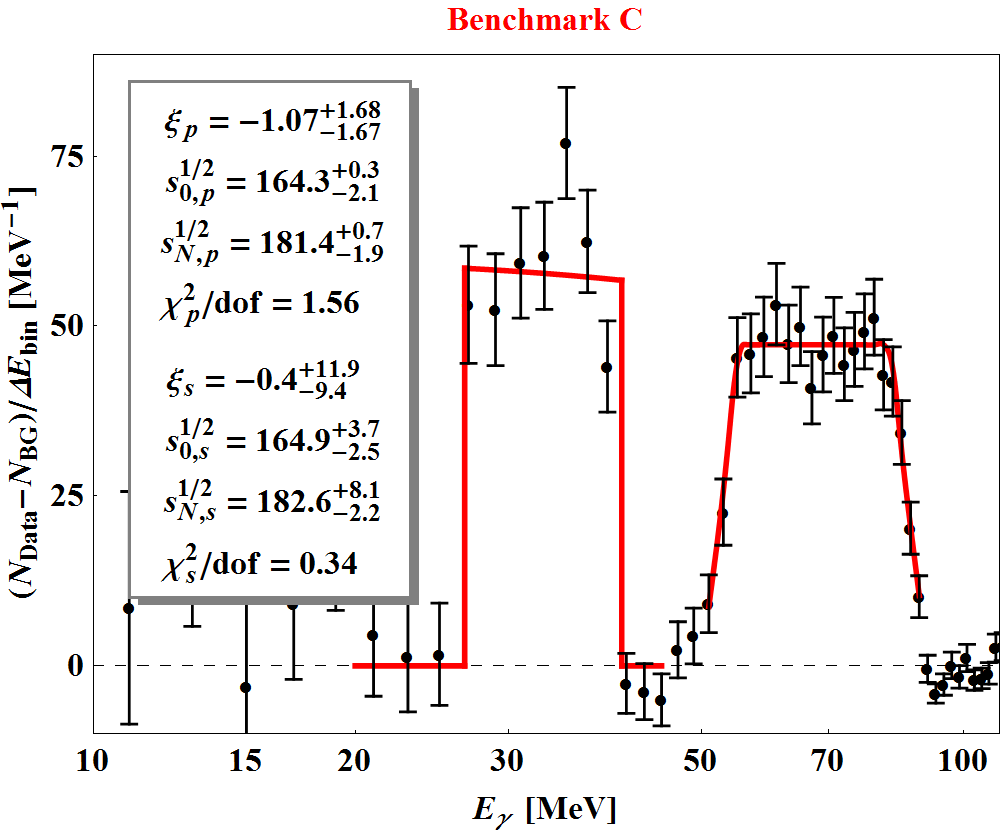} \hskip 1.0cm
\includegraphics[width=0.425\textwidth, keepaspectratio]{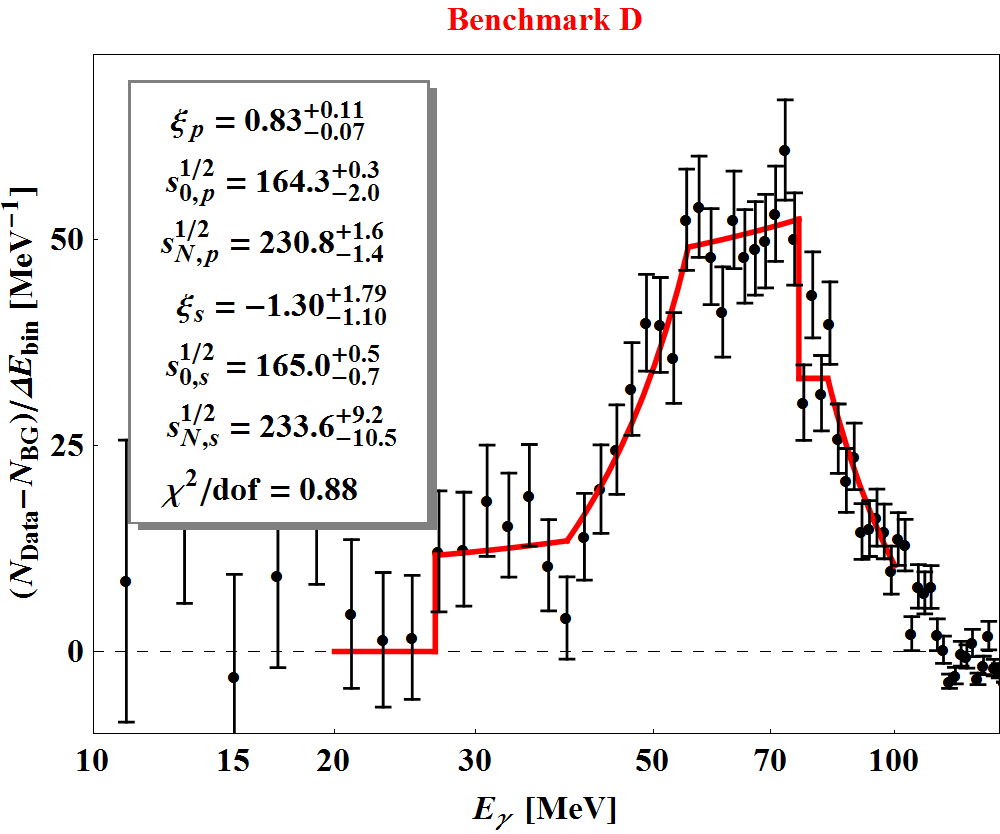}
\caption{Sample photon-energy spectra (black dots with corresponding statistical error bars) 
for Benchmarks~A (upper left panel), B (upper right panel), C (lower left panel), 
and D (lower right panel) after background subtraction, along with the corresponding 
best fits for the primary  and secondary photon spectra (solid red curves).   
For each benchmark, the numbers of background and signal events are taken to be 
$N_B = 2.32\times10^5$ and $N_S = 2.41\times10^3$, as discussed in the text.  
Note that we plot the quotient $(N_{\rm Data}-N_{\rm BG})/\Delta E_{\rm bin}$ on the vertical axis
(where the numerator tabulates the signal counts within each bin and the denominator indicates the corresponding
bin size), as this quotient is invariant under changes in the specific choice of bin size 
when the bin size is sufficiently small.
The corresponding error bars, by contrast, depend on bin size, 
and we have chosen $\Delta E_{\rm bin}=2~{\rm MeV}$ for the curves in these plots. 
The best-fit parameters are also indicated within each panel, 
along with the corresponding goodness-of-fit 
$\chi^2$ per degree of freedom, while  the upper and lower uncertainties quoted for each best-fit parameter 
indicate the limits of the corresponding range within which $\chi^2$ varies by less than one unit.
Note that the fits performed here are {\it unconstrained}\/, in the sense
that the primary  and secondary photon spectra are fit independently.   These fits thus provide a test of the extent to which the correlations between these two spectra can be discerned from data.}
\label{fig:fitBMA}
\end{figure*}

The results of our analysis are as follows.
For each of the benchmarks listed in Table~\ref{tab:case},
our corresponding simulated data set is shown in Fig.~\ref{fig:fitBMA} (black dots with error bars).
Specifically, these dots represent the residual
populations of events $\mathcal{N}_i^{\mathrm{Sig}}$ in the relevant energy bins,
with error bars corresponding to statistical uncertainties.  
Also superimposed on these data sets are the results of parametric fits 
to the spectral features associated with primary and secondary photons
(solid red lines).
Recall that in these plots, the spectral features associated with the 
primary and secondary flux spectra are fit independently.
As discussed above, these are the fits which are designed to address the first question itemized above.

The results for Benchmark~A are shown in the upper left panel of Fig.~\ref{fig:fitBMA}.
For Benchmark~A, our value of $N_S$ translates into the result
\beq
     \Xi ~=~ 5.4 \times 10^{-9}~ {\rm cm}^{-2} \,{\rm s}^{-1}\, {\rm MeV}^{-1}~,
\label{inputval}
\eeq
which we take as our input value for this benchmark. 
We perform our fit to the primary photon spectrum for Benchmark~A within
the energy range $20~\mev \leq E_\gamma \leq 45~\mev$ --- indeed,  the region $E_{\gamma}<20~\mev$ is background-dominated, leaving the corresponding bin counts less reliable, given our signal statistics.
We find that the best-fit values for $\xi_p$ and $\sqrt{s_{N,p}}$
are those indicated in the upper left panel of Fig.~\ref{fig:fitBMA}.
It is immediately evident that these extracted values are consistent with the corresponding input values 
to within $1\sigma$. 
Note that no meaningful
information can be extracted for $\sqrt{s_{0,p}}$, as
large uncertainties in the
event counts in bins with $E_\gamma \lesssim 20~\mev$ completely obscure all meaningful information about the low-energy cutoff in the primary photon spectrum.  
Thus, a best-fit value for $\Xi$ is not available from the primary photon spectrum, as this would require 
the value for $\sqrt{s_{0,p}}$.

By contrast, a fit to the secondary photon spectrum for Benchmark~A provides far more reliable
information about the properties of the underlying DDM ensemble.  Performing such a fit
within the energy range $50~\mev \leq E_\gamma \leq 90~\mev$ where the residual bin counts are greater than $\sim 10$, we find the results shown in the upper left panel of Fig.~\ref{fig:fitBMA}.
Once again, each of these extracted values is in good agreement with the corresponding input value to within $1\sigma$.
Since we are able to meaningfully extract $\sqrt{s_0}$ for the secondary photon spectrum, we are also able to report 
the best-fit value for $\Xi$ in this case. 
We find that the best-fit value for the normalization parameter is
\begin{equation}
  \Psi_s ~=~ 4.8^{+275.9}_{-4.8} \times 10^{-11} ~  {\rm cm}^{-2} \, {\rm s}^{-1} \, {\rm MeV}^{-1-\xi}~,
\end{equation}
from which we obtain the value of $\Xi$ for the secondary photon spectrum:
\bea
   \Xi_s ~=~  3.0^{+169.0}_{-12.3}\times 10^{-9} ~ {\rm cm}^{-2}\, {\rm s}^{-1}\, {\rm MeV}^{-1}~.
\eea
Although this extracted value is consistent with the input value in Eq.~(\ref{inputval})
to within $1\sigma$, the corresponding uncertainty is too large to infer any useful information.

Thus, for Benchmark~A, we conclude that it is difficult to obtain meaningful information 
concerning the normalization parameter $\Xi$ from either the primary  or secondary photon spectrum.   
By contrast, we see that reasonable estimates of the parameters which govern the 
 {\it shapes}\/ of the primary and secondary photon spectra 
can indeed potentially be obtained from future gamma-ray detectors.
Moreover, 
the fact that the values of DDM parameters such as $\sqrt{s_N}$ and $\xi$
extracted from the primary photon spectrum
match those extracted from the secondary photon spectrum
implies that we can indeed perform a successful test of the underlying correlations between these two spectra, 
and indicates that our primary and secondary spectra together contain 
consistent information regarding the underlying DDM model.

We now turn to Benchmark~B, for which results are shown in the
upper right panel of Fig.~\ref{fig:fitBMA}.
For this benchmark, the signal events are produced within the energy range $135~\mev < \sqrt{s} < 231~\mev$,
requiring us to take 
\beq
     \Xi ~=~  2.3\times 10^{-9} ~{\rm cm}^{-2} \, {\rm s}^{-1} \, {\rm MeV}^{-1}~
\eeq
as an input value.
Since the primary and secondary photon spectra overlap significantly for this
benchmark, we perform a combined fit to both features in the manner discussed
above, taking our  fitting range to be $20~\mev \leq E_{\gamma} \leq 100~\mev$.  We then obtain 
the best-fit values for the shape parameters
given in Fig.~\ref{fig:fitBMA}.
Once again, we observe
that the parameters $\xi$ and $\sqrt{s_N}$ extracted for both spectra
agree reasonably well with each other, thus providing a rough test of their correlation.
Moreover, each of the extracted shape parameters listed in the upper right panel of Fig.~\ref{fig:fitBMA}
is consistent with the corresponding input value to within $(1-2)\sigma$.

Thus, for Benchmark~B, we conclude that our fitting procedure yields reasonable estimates for the shape parameters which characterize the photon spectrum associated with our DDM ensemble.
The best-fit value of $\Psi$, by contrast, comes with large uncertainties.
Indeed, we shall find that this is a characteristic of all of the benchmarks we shall be examining.
We shall therefore refrain from 
quoting further best-fit values for $\Psi$ and $\Xi$ in what follows.
However, we stress that in all cases this is strictly only an artifact of the parametrization
and does not represent a corresponding uncertainty
in actual signal flux or in the number of signal events (the uncertainty for which is indeed small).

Note that Benchmark~B provides a better handle for measuring the DDM scaling parameter 
$\xi$ accurately, especially when
compared to $\xi_p$ in Benchmark~A.~
This is ultimately because more signal events in Benchmark~B are populated in the higher-energy regime where the background contribution is relatively small.
Therefore, even after background subtraction, the residual spectrum for Benchmark~B better preserves the original shape information than it does for Benchmark~A.

For the remaining two benchmarks, even the primary photon spectrum has a reasonable sensitivity to $\sqrt{s_0}$ because it starts from $E_{\gamma}>20~\mev$ where uncertainties in the event counts in bins are fairly decent.  
Our results for Benchmark~C are shown in the lower left panel of Fig.~\ref{fig:fitBMA}. 
The signal events are generated with $164~\mev < \sqrt{s} < 180~\mev$, from which we find 
that 
\beq
     \Xi ~=~  1.6\times 10^{-8} ~{\rm cm}^{-2} \, {\rm s}^{-1}\, {\rm MeV}^{-1}~.
\eeq
As with Benchmark~A, the two photon spectra are well separated, and thus two individual fits are possible. 
We adopt the same energy ranges as for Benchmark~A, namely 
$20~\mev \leq E_{\gamma} \leq 45~\mev$ and $50~\mev \leq E_{\gamma} \leq 90~\mev$, respectively,
for our fits to the primary and secondary photon spectra,
and obtain the best-fit results 
for the shape parameters 
as shown in the figure.
The parameters for the primary  and secondary photon spectra are generally consistent with each other,
thus indicating the possibility of testing correlations between them,
and they are also in a good agreement with the corresponding input values to within $(1-2)\sigma$. 
It turns out that the overall shape of the secondary photon spectrum does not change much for this benchmark, 
even with substantial variations of the scaling parameter. 

\begin{figure*}[t]
\includegraphics[width=0.425\textwidth, keepaspectratio]{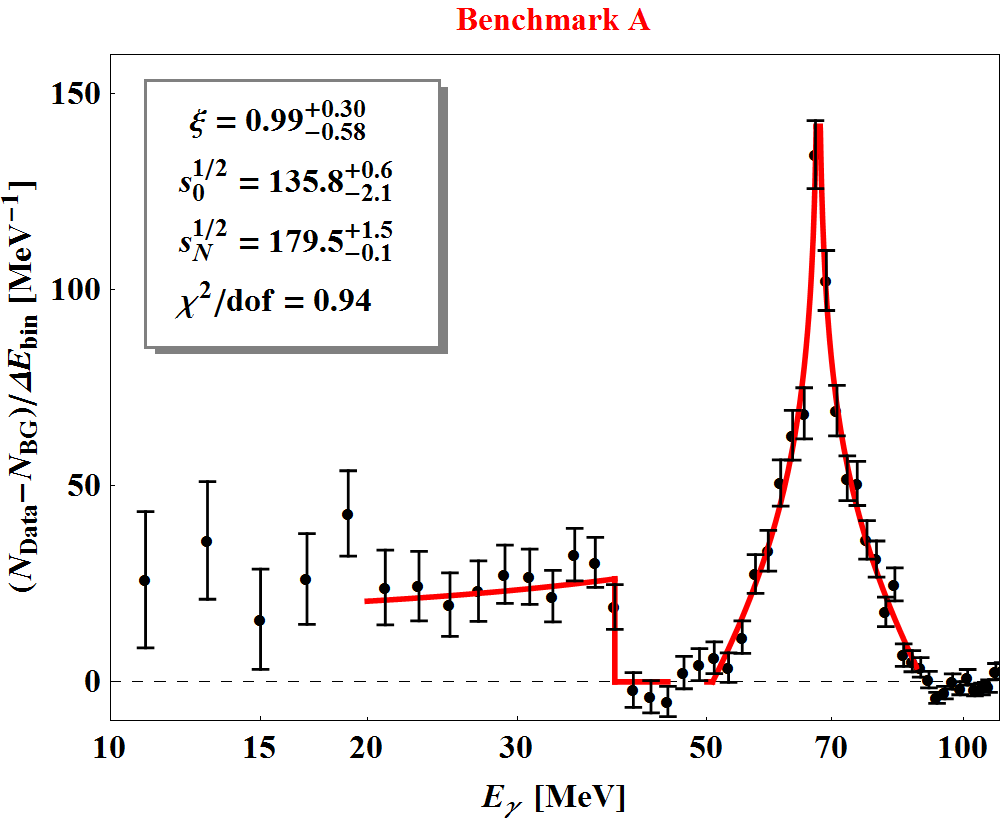} \hskip 1.0cm
\includegraphics[width=0.425\textwidth, keepaspectratio]{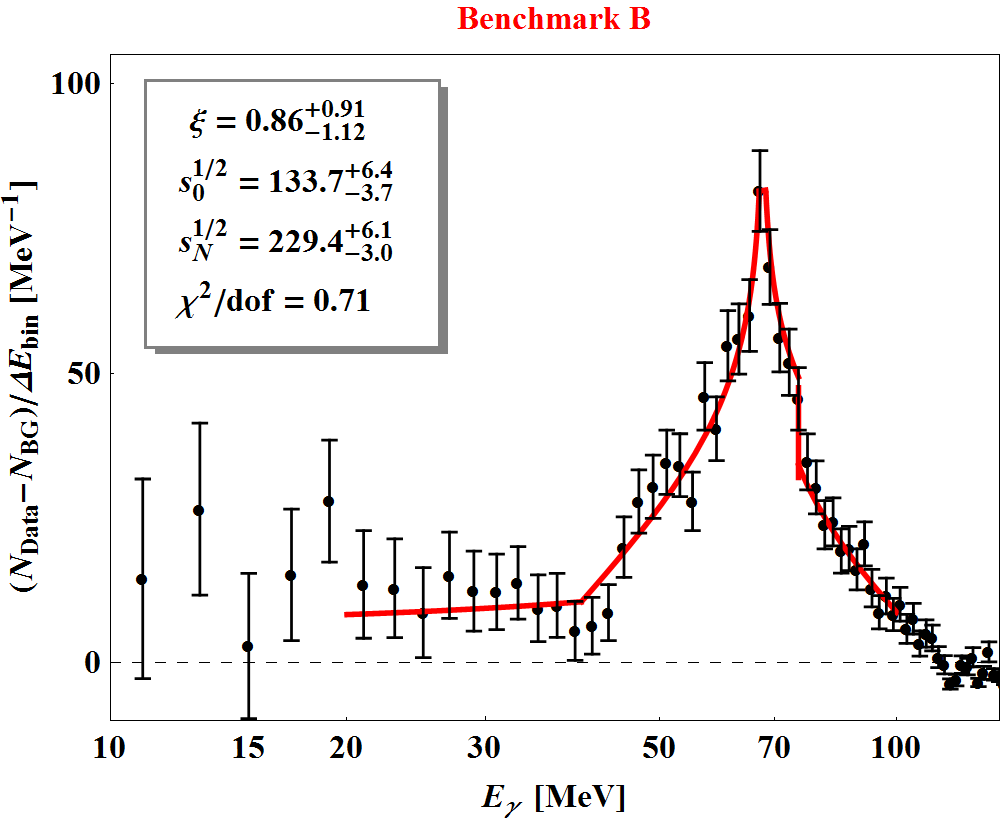} \vskip 0.4cm
\includegraphics[width=0.425\textwidth, keepaspectratio]{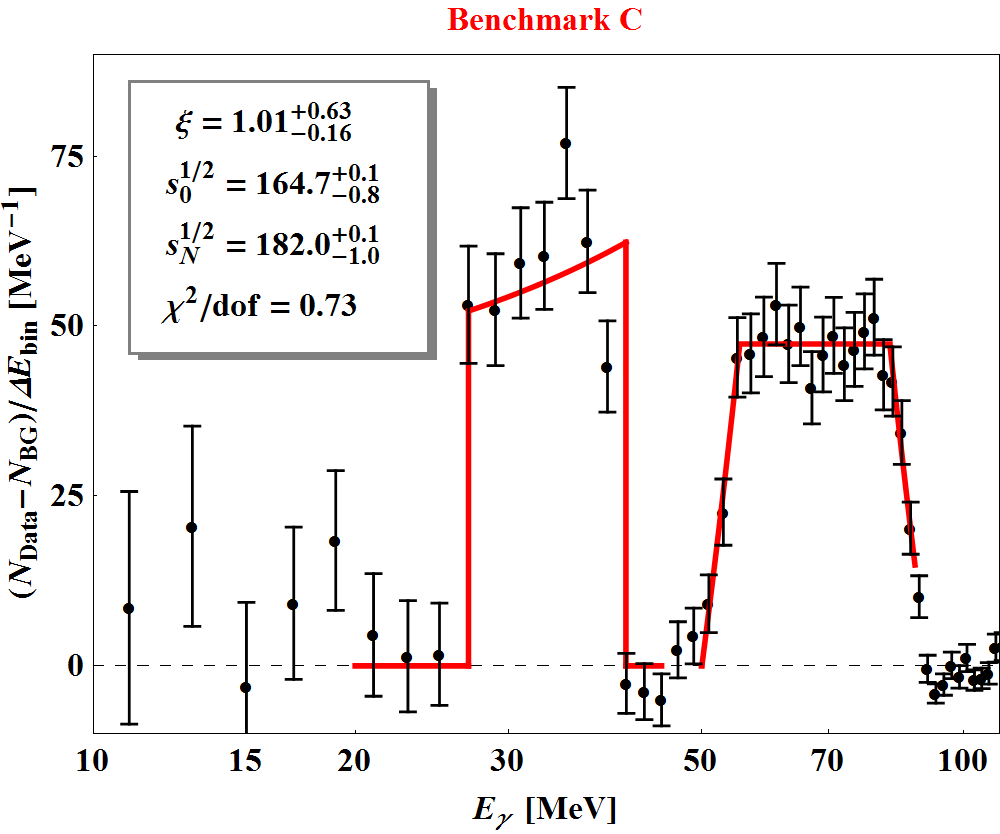} \hskip 1.0cm
\includegraphics[width=0.425\textwidth, keepaspectratio]{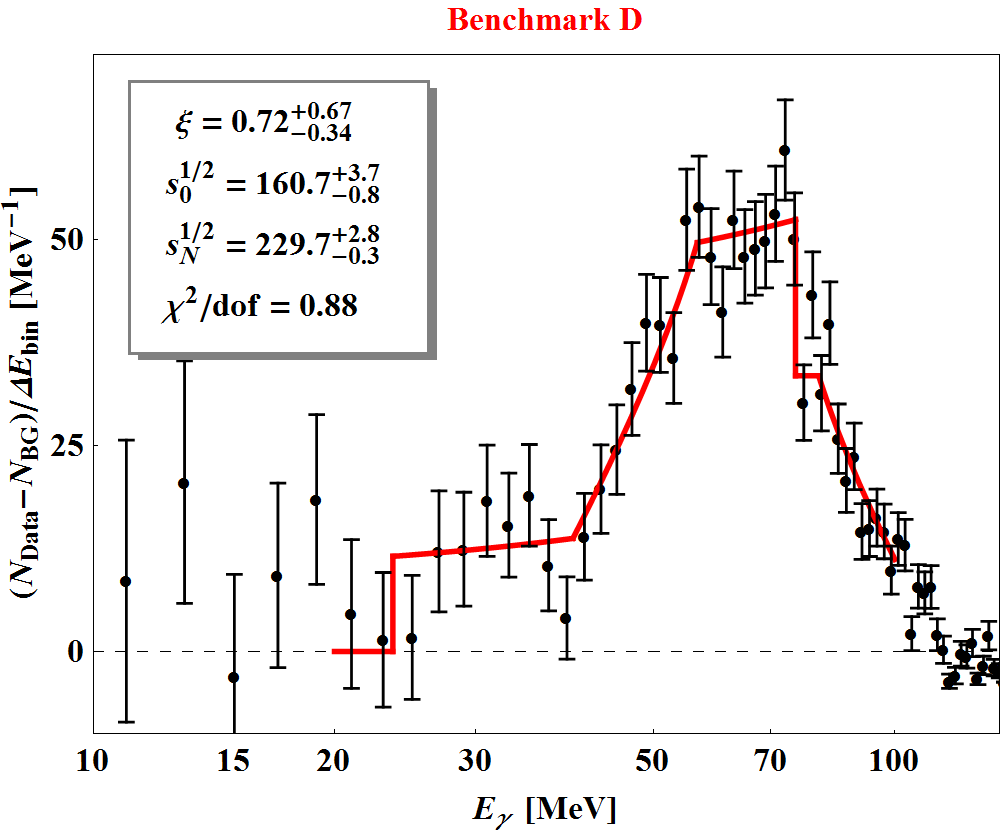}
\caption{Same as Fig.~\protect\ref{fig:fitBMA}, except that we now perform {\it constrained}\/ fits in which the 
primary  and secondary photon spectra are assumed to be correlated.
A comparison with the results of Fig.~\protect\ref{fig:fitBMA} demonstrates 
that the assumption of such correlations can significantly enhance our ability 
to accurately extract the underlying DDM parameters governing the dark sector. }
\label{fig:finalfignew}
\end{figure*}

Results for Benchmark~D are shown in the lower right panel of Fig.~\ref{fig:fitBMA}. 
The signal events are generated with $164~\mev < \sqrt{s} < 230~\mev$, from which we find
\beq
     \Xi ~=~  3.7\times 10^{-9} ~{\rm cm}^{-2}\, {\rm s}^{-1}\, {\rm MeV}^{-1}~.
\eeq
We then perform a single combined fit to both photon spectra,
again adopting the same fitting range $20~\mev \leq E_{\gamma} \leq 100~\mev$  
as for Benchmark~B.~
The best-fit values for all shape parameters are listed in Fig.~\ref{fig:fitBMA}.
We can easily see that the parameters measured from both spectral features are 
consistent with each other, as expected.  
The extracted values also all agree with the corresponding input values to within $(1-2)\sigma$. 

In general, scanning the results in Fig.~\ref{fig:fitBMA} for all four benchmarks 
simultaneously, we see that our best-fit results for $\sqrt{s_0}$ and $\sqrt{s_N}$
are generally quite accurate.   Unfortunately, we also observe that these fits
generally do a poor job of extracting the true underlying values of the 
DDM scaling parameter $\xi$.
While certain benchmarks (such as Benchmark~B) lead to 
relatively accurate best-fit values for $\xi$, particularly for the primary photon spectrum, 
these predictions become significantly worse for those benchmarks (such as
Benchmarks~A and C) in which the spectral features associated with 
the primary and secondary photons are relatively well-separated in energy, 
with minimal overlap.
The case of Benchmark~C is particularly poor, with {\it negative}\/ central values of $\xi$
extracted from both the primary and secondary spectra!
Indeed, the negative central value for $\xi_p$ is reflected in the negative slope of the red best-fit line
along the primary plateau in the lower left panel of Fig.~\ref{fig:fitBMA}.
 
All of the fits performed thus far treat our primary  and secondary photon 
spectra independently.  As discussed above, they are therefore suitable for addressing 
the first bulleted question at the beginning of this section
concerning the extent to which correlations between the two photon fluxes might
be discernible in realistic data samples.
However, in order to address 
the second of our bulleted questions, we need to assume the existence of such correlations and 
perform constrained fits to both spectra simultaneously.
Indeed, it is only by performing such constrained fits and comparing the results thus obtained
with those of the unconstrained fits we have already perfomed that we can determine the extent to which
these correlations enhance our ability to extract the underlying DDM parameters from data.

The results of such constrained fits are shown in Fig.~\ref{fig:finalfignew}.
Upon comparison with the corresponding results in Fig.~\ref{fig:fitBMA},
we immediately see that while our extracted best-fit values of $\sqrt{s_0}$ and $\sqrt{s_N}$ continue
to be as accurate as they were before,
our extracted best-fit values for the DDM scaling parameter $\xi$ are significantly improved.
Indeed, in all cases the true value 
$\xi=1$ is within the errors quoted.
The case of Benchmark~C is particularly noteworthy.
Where previously our unconstrained fits had yielded negative values for both $\xi_p$ and $\xi_s$,
the simple act of changing to a constrained fit has pushed the corresponding best-fit 
result to a central value $\xi= 1.01$, which is remarkably close to the true value! 
In general, we see that it is Benchmarks~A and C --- \ie, benchmarks in which our two spectral features are
well separated in energy --- for which the switch from
an unconstrained fit to constrained fit
produces the greatest improvement.
It is thus these benchmarks for which the assumption of a correlation between the primary
and secondary photon spectra is of greatest value.
Indeed, as evident from Fig.~\ref{fig:finalfignew},
the assumption of a correlation between the primary and secondary flux spectra
leads to a signficant improvement in our ability to extract the underlying DDM parameters 
regardless of the particular benchmark under study.

Of course, our comparison between the fits in Fig.~\ref{fig:fitBMA} and 
those in Fig.~\ref{fig:finalfignew}
amounts to analyzing the results of only a single pseudo-experiment.   In principle, one could rerun this
experiment with many different random data sets, and repeat this analysis in each case.
However, we shall refrain from this exercise because
the main points that we have aimed to demonstrate are already evident.
Indeed, the results illustrated in 
Figs.~\ref{fig:fitBMA} and \ref{fig:finalfignew}
prove to be both typical and robust.

We conclude, then, that it will indeed be possible  
to extract evidence of a correlation
between primary and secondary photon 
spectra at future gamma-ray facilities. 
Moreover, we see that 
the assumption of such a correlation
will indeed significantly sharpen our ability to extract 
the corresponding underlying dark-sector parameters.
Thus, through this correlation, we see that our ability to indirectly probe
the physics of the dark sector through emitted 
gamma-rays can be greatly enhanced.


\section{Conclusions and Outlook\label{sec:conclusions}}


In this paper, we have identified an unambiguous indirect-detection signature of
Dynamical Dark Matter which arises in cases in which the constituents of the
DDM ensemble annihilate or decay primarily into a final state involving a primary
photon and a neutral pion, the latter subsequently decaying into a pair of secondary photons.
When the mass gap between DDM constituents is sufficiently small that particle detectors 
are unable to resolve the contributions of individual constituents in the photon energy spectra,
this signature involves a pair of characteristic {\it continuum}\/ features in the
gamma-ray spectrum in the $\mathcal{O}(1 - 100)~{\rm MeV}$ range --- one feature associated
with the primary photons, and the other feature associated with the secondary photons.  
Since the spectral shapes
of these two features are correlated, a comparison between the information extracted
from the two continuum features provides a powerful consistency check that
they indeed have a common origin in terms of an underlying DDM ensemble.
We have examined the prospects for observing a signal of this sort at the
next generation of MeV-range gamma-ray telescopes and investigated the extent
to which the parameters which govern the DDM ensemble can be extracted from
spectral data once such a signal is unambiguously identified.
As we have seen, it should be possible not only to extract evidence of this correlation
in future photon spectral data, but also to exploit this correlation in  
order to significantly enhance our ability to extract these underlying DDM parameters.

A few comments are in order.  First, in order to maintain maximum generality,
we emphasize that
we have estimated both the discovery
reach and the potential for measuring the DDM model parameters at the next generation
of gamma-ray detectors by defining a simplified, hypothetical detector whose attributes
have been chosen not to be identical to those of any particular such instrument,
but rather to be representative of this class of experiments in general.  
However, for a realistic
detector, the corresponding analysis would typically involve additional subtleties and
complications.  For example, the energy resolution for such a detector is typically
not described by a Gaussian smearing function with a constant value of $\epsilon$.
Moreover, the effective area for a realistic detector is typically not
independent of photon energy throughout the range of $E_\gamma$ to which the
instrument is sensitive.

In addition to these experimental simplifications, there are also a number of theoretical
approximations which we have employed in our analysis.
For example, we have taken the branching fraction for the
annihilation/decay of all ensemble constituents to the $\gamma \pi^0$
final state to be effectively unity.  However, there are situations in which this is
not necessarily true for the lightest ensemble constituents.  The reason is that
a fundamental interaction between the $\phi_n$ and SM quarks of the sort
which leads to dark-matter annihilation/decay to $\gamma \pi^0$ also
generically leads to annihilation/decay to $e^+ e^-$ and/or $\mu^+ \mu^-$,
via loop-level processes involving a virtual photon.  The branching fraction into
such leptonic final states is typically negligible for most of the $\phi_n$.
However, it can become significant for processes in which the CM
energy is only slightly above the kinematic threshold $\sqrt{s}_n  \approx m_{\pi^0}$
for $\gamma \pi^0$ production.  As a result, the sharpness of the peak in the
secondary photon spectrum at $E_\gamma \approx m_{\pi^0}/2$ depends both on
$\sqrt{s_0}$ and on the energy resolution of the detector.  Incorporating these
considerations into a more detailed analysis would inevitably lead to a modification
of our quantitative results in DDM scenarios of this sort.

On a related note, we remark that our focus 
in this paper
has been on the case in which that the dominant signal contribution
to the photon flux arises from ensemble constituents whose
CM energy for annihilation/decay lies within
the range $m_{\pi^0} \leq \sqrt{s} \leq 2m_{\pi^\pm}$.  However, it
is also useful to consider how our results would be affected if
non-trivial contributions to the photon flux were also to arise from
constituents with $\sqrt{s}$ outside this range, and thus from
photoproduction processes with different kinematics.  For example,
it is important to examine whether such contributions might obscure the
spectral features which we have discussed in this paper.

We begin by considering the contribution from constituents with
$\sqrt{s}$ slightly above the $2m_{\pi^\pm}$ threshold, for which the dominant
$C$-odd final state will be $\pi^+ \pi^-$.  The principal contribution from the
photon flux in this case arises from final-state radiation.  Photons produced in
this way tend to be quite soft, and as a result, any contamination of our
signal spectrum from such photons would primarily affect the region where
$E_\gamma$ is low and statistical power is already poor.  By contrast,
for the $\gamma \pi^0$ final state which has been the focus of our paper,
at least one of the two salient spectral features always appears at a relatively high
energy.  For constituents with even larger $\sqrt{s}$, for which final states involving
three or more pions are accessible, the shape of the resulting photon spectrum
becomes highly model-dependent.  However, one generally expects these spectra to
be relatively smooth and featureless over the range of $E_\gamma$ relevant for
our analysis.

Now let us turn to the contribution from constituents with $\sqrt{s} < m_{\pi^0}$.
For $\sqrt{s}$ in this regime, the dominant contribution to the photon flux arises
from the final state $3\gamma$, and from final-state radiation produced in conjunction with
the final state $e^+ e^-$.  The former contribution is associated with processes
involving an off-shell $\pi^0$, while the latter is associated with processes involving an off-shell
photon attached to a quark loop.  Photons produced in conjunction with the $e^+ e^-$
final state will once again be quite soft and consequently have little impact on our
results.  By contrast, the contribution from the $3\gamma$ final state could potentially
distort the shape of the secondary-photon spectrum at energies slightly below its peak.

In this paper, we have applied our analysis to a DDM ensemble in which the photon flux scales
with the center-of-mass energy as a power law.  Examples of explicit DDM models in which
such behavior is exhibited include those in Refs.~\cite{DDM1,DDM2}.  
Indeed, as noted above Eq.~(\ref{eq:flux}), scaling relations
of this form
tend to emerge naturally for a variety of
theoretical structures
underlying these ensembles.
However, there do exist DDM constructions in which such scaling relations are given not by simple power 
laws but by other functional forms~\cite{RandomMatrixDDM,HagedornDDM}.
These situations can nevertheless be addressed in a manner similar to that which we have 
employed in this paper.
In general, the photon flux is determined by the scaling of the abundance and
annihilation/decay rate as in Eq.~(\ref{eq:PhintotExp}).   Thus, for any other DDM construction,
one can similarly determine the primary and secondary photon fluxes.  In fact,
it is not even necessary for the dark sector to constitute a DDM ensemble at all. 
Even if the dark sector
consists of multiple particles whose lifetimes and abundances are not determined by any
unified organizing principle, those lifetimes and annihilation/decay rates completely determine the primary and
secondary differential photon fluxes.  

The lynchpin of this paper has been the correlation between the spectral shapes of
the primary and secondary photon fluxes.
Fortunately, this correlation is robust and survives even if the
dark sector lacks a unified organizing principle.
To see this most directly, we recall that
each dark-sector constituent 
of a given mass makes only a single monochromatic contribution to the primary photon flux.
Thus, the relation between the primary flux and the underlying
dark-sector component masses is easily invertible:  if the primary flux is known, 
then one can easily determine the spectrum of particles and annihilation/decay
rates which generated that primary spectrum.  This in turn then provides a prediction for the secondary photon flux.  

Using the primary photon flux to predict the secondary photon flux is a strategy that 
is likely to be most useful in the case where $\sqrt{s_N} < \sqrt{2} \mp$,  for which the primary and secondary photon features can be cleanly separated.  Indeed, after subtracting the
estimated background from the data in the region of the primary feature, the residuals constitute
a measurement of the primary photon flux, up to statistical fluctuations and the smearing due to the
energy resolution.  One could then use this primary photon flux to generate a prediction for the secondary
flux, and test the goodness of fit for this prediction to the actual data in the region of the secondary
feature.  
However, since the determination of the primary
photon flux is distorted by the statistical fluctuations and the effects of a finite energy resolution,
the implementation of this strategy is likely to be non-trivial. 
This would therefore be an interesting direction for future study.

In cases for which  $\sqrt{s_N} > \sqrt{2} \mp$,  by contrast, the primary and secondary photon features are expected to overlap significantly.   As a result, it may be more problematic to cleanly separate them.
Despite this fact, we have already seen that these two features remain correlated and in the case of a DDM ensemble we have seen that this correlation can significantly enhance our ability to extract the underlying DDM parameters --- even 
when these features overlap significantly.
In general, however, performing an {\it a priori}\/ separation of the primary and secondary photon features 
will undoubtedly be a more complicated task in the cases where these features overlap.
One useful tool in this regard may be to exploit the so-called ``log-symmetry'' of the secondary photon flux --- \ie, the invariance of this flux under the energy-inversion symmetry $E\to \mptwo/4E$, as discussed in the Appendix.
Any contributions to the total flux which violate this symmetry are necessarily those from the primary photons.

Finally, in closing, we remark that correlations between continuum features
which arise in the gamma-ray spectra of annihilating/decaying DDM ensembles
arise not only for the $\gamma \pi^0$ final state
which has been the focus of this paper, but for other final states as well.
For example, in DDM scenarios in which each of the ensemble constituents
can annihilate into both $\gamma\gamma$ and $\gamma Z$,
similar correlations between the shapes of the two resulting
spectral features can likewise be exploited in order to corroborate the DDM
origin of the excess and to extract information about the parameters
governing the underlying ensemble.  Thus, such correlations could likewise be used
in order to extract information about this alternative class of DDM ensembles.


\begin{acknowledgments}


We would like to thank Kaustubh Agashe and Graciela Gelmini for useful discussions.
We would also like to acknowledge the hospitality
of the Center for Theoretical Underground Physics and Related Areas (CETUP$^\ast$),
where this work was initiated
during the 2015 Summer Program; 
DK, JK, JCP, and BT would also like to thank CETUP$^\ast$ for partial support during this program.
KB and JK are supported in part by the National Science Foundation under CAREER Grant PHY-1250573,
while KRD is supported in part by the Department of Energy under Grant DE-FG02-13ER41976
and by the National Science Foundation through its employee IR/D program.
DK is supported in part by the U.S.\ Department of Energy under Grant DE-SC0010296,
and JCP is supported in part by the Basic Science Research Program through the 
National Research Foundation of Korea funded by the Ministry of Education (NRF-2013R1A1A2061561).
BT is supported in part by an internal research award from Reed College. 
The opinions and conclusions expressed herein are those of the authors, and do not
represent any funding agencies.

\end{acknowledgments}


\appendix
\section{~~The Line and the Box:\\   Decay Kinematics for the Process $\phi_n\to \gamma\pi^0 \to \gamma \gamma \gamma$}

This appendix is dedicated to a quick review of the 
decay kinematics~\cite{Stecker,Agashe:2012bn,Boddy:2015efa}
associated with the primary decay process $\phi_n\to \gamma \pi^0$, 
followed by the secondary decay process $\pi^0\to \gamma \gamma$.
Our goal is to calculate the spectrum of energies of the photons produced through these processes,
as measured in the lab (detector) frame,
assuming that our initial ensemble constituent $\phi_n$ with mass $m_n$ decays from rest in this frame.

Understanding the primary decay process $\phi_n\to \gamma\pi^0$ is relatively straightforward.
With $\phi_n$ taken to be at rest at the time of its decay, 
conservation of energy and momentum  
immediately lead to the two constraint equations
\begin{eqnarray}
         m_n &=& E_\gamma^{(1)} + \mp \gamma_\pi \nonumber\\
         E^{(1)}_\gamma  &=& \mp \gamma_\pi \beta_\pi = \mp \gamma_\pi \sqrt{1-1/\gamma_\pi^2}~
\end{eqnarray}
where $E_\gamma^{(1)}$ is the energy of the primary photon and 
where ($\gamma_\pi$, $\beta_\pi$) denote the boost factor and corresponding velocity of the emitted on-shell pion.
Solving these equations, we find that the energy $E_\gamma^{(1)}$ of the primary photon is given by
\begin{equation}
       E_\gamma^{(1)} ~=~ \frac{m_n^2 - \mptwo}{2m_n} 
\label{primaryenergy}
\end{equation}
while $(\gamma_\pi,\beta_\pi)$ 
are given by
\beq
  \gamma_\pi ~=~ \frac{m_n^2 + \mptwo}{2 m_n \mp} ~,~~~~~ \beta_\pi = \frac{m_n^2 - \mptwo}{m_n^2 + \mptwo}~.~~~~
\eeq
Thus the lab-frame energy of the emitted pion is given by
\beq
      E_\pi ~=~  \mp \gamma_\pi ~=~ \frac{m_n^2 + \mptwo}{2 m_n}~.
\eeq

The second step is to determine the energies of the secondary photons.
In the rest frame of the emitted pion, these energies are nothing but $\mp/2$.
However our goal is to determine these energies as measured in the lab frame.
To do this, we need to account for the boost of the emitted pion.
Let us assume that one of the secondary photons is emitted at an angle $\theta$, as measured
in the rest frame of the pion,  relative to the boost direction of the pion.
We then find that lab-frame energy of this secondary photon is given by
\begin{eqnarray}
   E^{(2)}_\gamma &=&  \frac{\mp}{2}  \gamma_\pi \left( 1 + \beta_\pi \cos\theta\right) \nonumber\\
    &=&  \frac{ 1}{4 m_n } 
             \bigg[ m_n^2 +\mptwo +  (m_n^2 -\mptwo) \cos\theta\bigg]~.~\nonumber\\
\label{secondaryenergy}
\end{eqnarray}
The lab-frame energy of the other secondary photon is given by the same expression, but with
$\theta\to \theta+\pi$, or $\cos\theta\to -\cos\theta$.

The interpretation of these results is clear.
When many such $\phi_n$ decays occur, 
the primary photons always have the energy $E_\gamma^{(1)}$ given in Eq.~(\ref{primaryenergy}).
They are thus mono-chromatic, forming a spectral line (\ie, occupying a discrete point 
in energy space).
By contrast, the secondary photons can populate any of the energies given in Eq.~(\ref{secondaryenergy}),
depending on the angle $\theta$.
Since the probability distribution for photon emission is isotropic in the
rest frame of the pion, all values of $\cos\theta$ are sampled with equal probability.
As a result, with enough decays, the secondary photons fill out a spectral ``box'' in energy
space.   This spectral box stretches over the range
$[ \mptwo/2m_n, m_n/2]$
and is centered
at $E_\gamma= (m_n^2 + \mptwo)/4m_n$
with width $\Delta E_\gamma = (m_n^2 - \mptwo)/2m_n$.

Interestingly, it turns out that the energy of the line always happens to be 
equal to the width of the box!
For $m_n < \sqrt{2}\mp$, the line is to the left of the box,
while for $m_n > \sqrt{2} \mp$, the line is {\it inside}\/ the box.
As $m_n\to \infty$, the line approaches the right edge of the box but never passes beyond it.

It is worth noting that there is only one value for the energy which {\it always}\/ finds itself within this box,
regardless of the
width of the box (\ie, regardless of the boost of the pion or the value of $m_n$):
this is $E_\gamma=\mp/2$, corresponding to the energy of the secondary photons
in the pion rest frame~\cite{Stecker,Agashe:2012bn}.
This is indeed nothing but the location of the line to which the box collapses as $m_n\to \mp$.
This is also the {\it geometric mean}\/ of the energies encompassed within the box.
Indeed, since the box is otherwise flat as a function of the energy, 
the energy spectrum of the secondary photons is actually ``log-symmetric'' [\ie,
invariant under the mapping $E\to \mptwo/4E$, or 
equivalently $y\to -y$ where $y\equiv \log(2E/\mp)$].
While these assertions are somewhat trivial for the spectrum corresponding to the secondary photons from
the decays of a single field $\phi_n$,
the fact that these features are independent of $m_n$ guarantees 
that they will be preserved even for the accumulated spectra  
of secondary photons emitted via the decays 
of {\it multiple}\/  $\phi_n$ with different masses $m_n$.
Indeed, this holds true {\it regardless}\/  of the particular structure of the underlying DDM ensemble
to which the $\phi_n$ belong.
Note that these assertions
form the centerpiece of Ref.~\cite{Agashe:2012bn}, where they were exploited 
in a collider-based context.

Our analysis above has focused on the kinematics of the decay process 
$\phi_n\to \gamma\pi^0 \to \gamma \gamma \gamma$.
However, in this paper we are also interested in the corresponding annihilation process
$\phi_n^\dagger \phi_n\to \gamma\pi^0 \to \gamma \gamma \gamma$.
Fortunately, given the analysis above,
it is not difficult to extract the corresponding results for the case of
annihilation rather than decay:
under the assumption that the $\phi_n$ are extremely non-relativistic with respect 
to the lab frame, the only required change in the above analysis is the global replacement $m_n \to 2m_n$.
Thus, given the definition of $\sqrt{s_n}$ in Eq.~(\ref{sndef}), 
we see that the replacement
$m_n\to \sqrt{s_n}$
everywhere in the above analysis
will generalize our results to apply
to $\phi_n$ annihilations   
as well as decays.
This is the procedure followed in the main text.


\end{document}